\documentclass[aps,pre,twocolumn,groupedaddress,superscriptaddress,amsmath,amssymb,amsfonts,floats,floatfix,balancelastpage,showpacs]{revtex4-1}

\usepackage{epsfig} 
\usepackage{psfrag}
\usepackage{braket}
\usepackage{color}
\usepackage[usenames,dvipsnames]{xcolor}
\usepackage{textcomp}
\usepackage{graphicx}
\usepackage{subfigure}
\usepackage{multirow}
\usepackage{latexsym}
\usepackage{url}

\usepackage[pdftex]{hyperref}

\hypersetup{
    pdfauthor={Qingyou Meng, Gregory M. Grason},
    pdftitle={Defects in Conformal Crystals},
    colorlinks=true,
    linkcolor=blue,
    citecolor=blue,
    filecolor=blue,
    urlcolor=magenta
}

\newcommand{\grad}{{\bf \nabla}}
\newcommand{\rv}{{\bf r}}

\graphicspath{{figures/}}   
\begin{document}

\title{Defects in Conformal Crystals: Discrete vs. Continuous Disclination Models}

\author{Qingyou Meng}
\affiliation{Department of Physics, University of Massachusetts, Amherst,
  Massachusetts 01003, USA}
\author{Gregory M. Grason}
\affiliation{Department of Polymer Science and Engineering, University of Massachusetts, Amherst, MA 01003, USA}

\begin{abstract}

We study the relationship between topological defect formation  and ground-state 2D packings in a model of repulsions in external confining potentials.  Specifically we consider screened 2D Coulombic repulsions, which conveniently parameterizes the effects of interaction range, but also serves as simple physical model of confined, parallel arrays of polyelectrolyte filaments or vortices in type-II superconductors.  The countervailing tendencies of repulsions and confinement to, respectively, spread and concentrate particle density leads to an energetic preference for non-uniform densities in the clusters. Ground states in such systmes have previously been modeled as {\it conformal crystals}, which are composed of locally equitriangular packings whose local areal densities exhibit long range gradients.  Here, we assess two theoretical models that connect the preference for non-uniform density to the formation of disclination defects, one of which assumes a continuum distributions of defect, while the second considers the quantized and localized nature of disclinations in hexagonal conformal crystals.  Comparing both theoretical descriptions to numerical simulations of discrete particles clusters, we show the influence of interaction range and confining potential on the topological charge, number and distribution of defects in ground states.  We show that treating disclinations as continuously distributable well-captures the number of topological defects in the ground state in the regime of long-range interactions, while as interactions become shorter range, it dramatically overpredicts the to growth in total defect charge.  Detailed analysis of the discretized defect theory suggests that that failure of the continuous defect theory in this limit can be attributed to the asymmetry in the preferred placement of positive vs. negative disclinations in the conformal crystal ground states, as well as a strongly asymmetric dependence of self-energy of disclinations on sign of topological charge.
 
\end{abstract}

\pacs{
  61.72.-y, 
  61.72.Bb,	
  61.72.Lk	
}

\maketitle

\section{\label{sec:introduction}Introduction}
Canonical forms of crystalline order arise as low-temperature phases of atoms, particles or molecules stabilized by short-ranged, cohesive  interactions.  Such interactions typically give rise to transitionally periodic ground states, and in such systems, topological defects appear only as excitations of a thermodynamically stable defect-free crystal~\cite{nelson_defects_2002}.  Low-temperature states formed by confined and mutually repulsive systems tend to break the prevailing paradigm of crystalline ground states, provided that either the interactions or the confining field acts over long range.  Examples of such systems include multi-vortex arrays of confined superconductors~\cite{Geim1997,Schweigert1998, Baelus2004, Cabral2004, Milosevic2003, Menezes2017}, one-component plasmas~\cite{bedanov_ordering_1994, koulakov_charging_1998, mughal_topological_2007, FORTOV2005}, confined polyelectrolyte bundles~\cite{Knobler2009, yao_topological_2013}, trapped, dipolar colloids ~\cite{Soni2018} and particles at liquid crystalline interfaces~\cite{Straube2019}.   In these systems, gradients in the pressure throughout the structure lead to a generic thermodynamic preference for {\it non-uniform} local density throughout the structure, which is generically incompatible with periodic lattice order of regular lattices.  

\begin{figure}[b!]
	\centering
	\includegraphics[width=0.45 \textwidth]{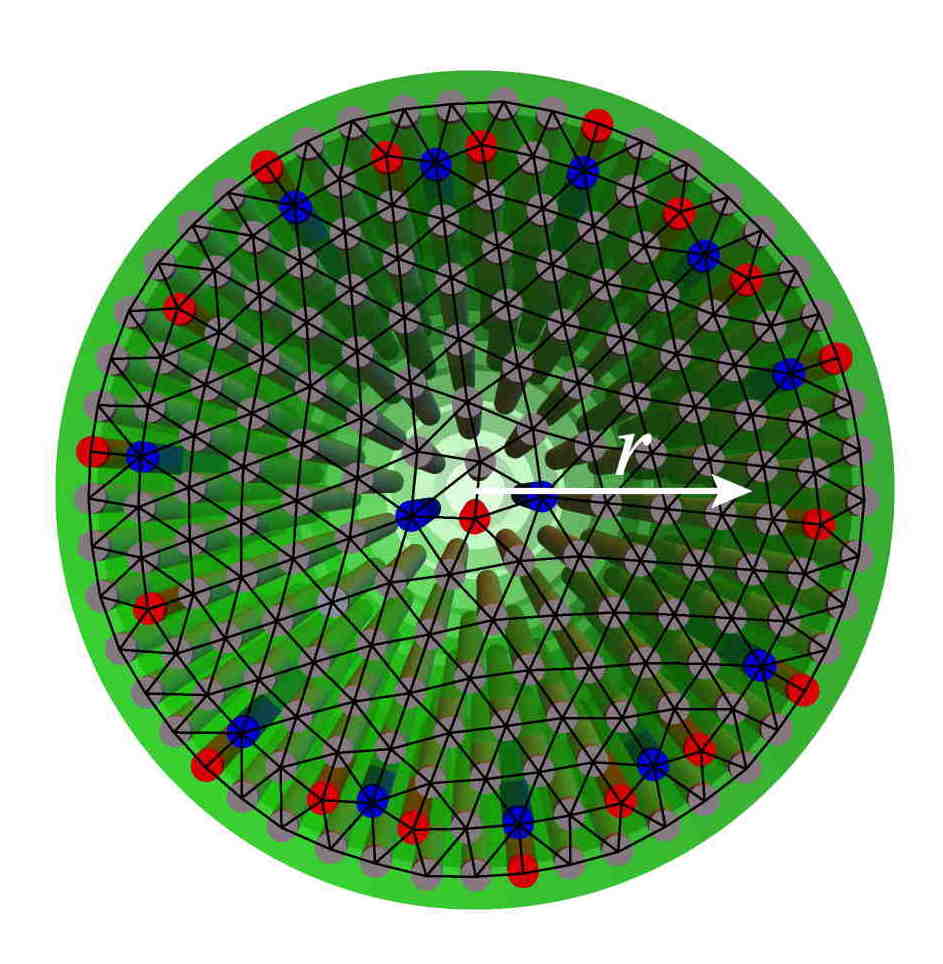}
	\caption{Conformal crystal ground state of $N=194$ parallel filaments, where semi-transparent green illustrates radially confining potential field.  Bond network (shown in black) highlights variable area density and predominance of local 6-fold symmetry of order, with defects shown as red and blue filaments, respectively, for 5- and 7-fold disclinations.  The 2D pattern was computed numerically for the 2D screened Coulomb repulsion model described below.}
	\label{fig:schematic}
\end{figure} 

For 2D ordered structures, the focus of this article, isotropic repulsion tends to favor locally equitriangular packing, in which six neighboring particles are distributed symmetrically around any given particle (see e.g. Fig. ~\ref{fig:schematic}).  At the same time, the energetic preferences for non-uniform local density then favor ground states where this locally triangular motif is isotropic dilated and contracted throughout the structure.  That is, such states can be modeled by conformal deformations of uniform triangular lattices, and hence, they are known as {\it conformal crystals}~\cite{Rothen_1993, Rothen_1996}.  Conformal crystals are characterized by unusual structural features, most notably patterns of long-range bending of lattice rows and associated excess densities of disclination defects~\cite{mughal_topological_2007}.  Disclinations are point-like rotational defects in the bond orientation, and in hexagonal crystals, they take the form of points that deviate from six-fold packing~\cite{seung_defects_1988}.  These defects are characterized by a topological charge $Q$, which characterizes the angular deficit of the bond-angle winding around the disclination, $2 \pi Q/6$ (i.e. $Q = +1$ and $-1$ correspond to five- and seven-fold defects).

The connection between non-uniformity of local spacing and these anomalous structure features has been the subject of numerous previous studies, nearly all of which connect these ground states to properties geometric properties of conformal maps~\cite{Rothen_1993, Rothen_1996, mughal_topological_2007, Soni2018, yao_topological_2013, Silva2020}.  Several of these studies have noted a connection between conformal maps and Gaussian curvature of 2D surfaces~\cite{yao_topological_2013, Soni2018}.  In short, gradients in local density can be characterized by a non-zero Riemannian curvature of the lattice.  Interpreting this curvature as the Gaussian curvature of a 2D surface, the ground state of the conformal crystal can be mapped onto this non-Euclidean surface in such as to preserve the locally-isotropic packing.   Building on the well-established theories that connect non-zero Gaussian curvature to topological defect formation in cohesive (i.e. uniform density) crystals~\cite{seung_defects_1988, Bowick2000, Vitelli2006, Bowick2009, Li2019}, it has been proposed~\cite{yao_topological_2013, Soni2018} that the ground-state defects patterns of conformal crystals should follow those that would form in their non-Euclidean analogs.  Namely, according to this picture, the local density of disclinations in conformal crystal ground state should tend to ``neutralize" the frustration imposed by Gaussian curvature, following a heuristic picture of defect screening of curvature-induced stresses in cohesive crystals~\cite{Irvine2010}.

\begin{figure*}[t!]
    \centering
    \includegraphics[width=0.8 \textwidth]{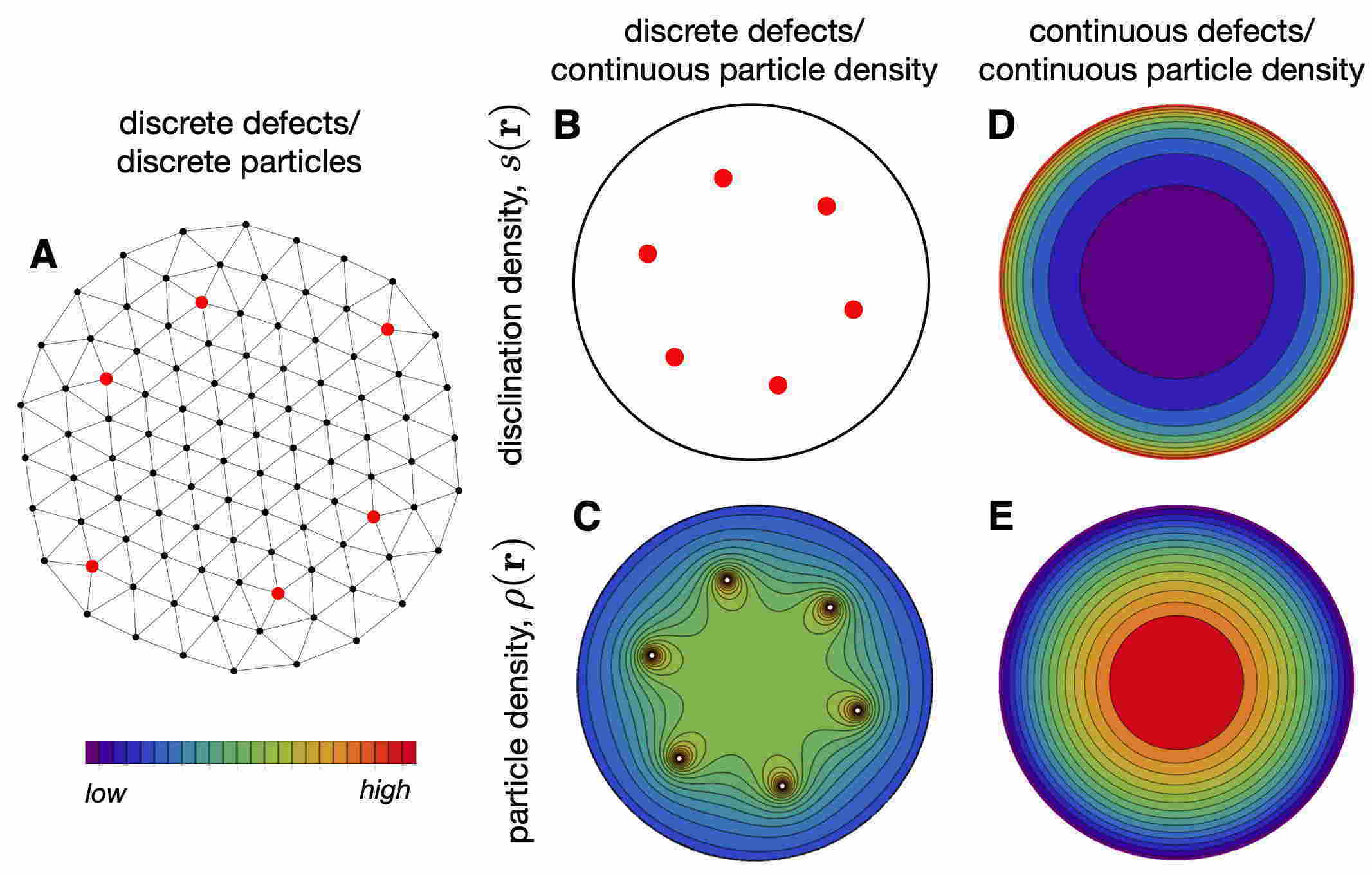}
    \caption{Schematic illustration of three theoretical frameworks for conformal crystal ground states:  (A), a discrete cluster of $N=100$ confined, repulsive particles with 6 discrete $Q=+1$ (i.e. 5-fold) disclinations shown in red.  (B)-(C), shows the ``Taylor'' picture in which defects are models as quantized and localize points, shows a red points in (B), and the corresponding inhomogeneous pattern of local particle density, as a continuous field (C).  (D)-(E) shows the ``Nye'' picture, in which disclination (D) and particle (E) distributions are continuously distributed fields.  Rendered patterns correspond to harmonic confinement in the limit of short range (2D Coulomb) interaction described below. }
    \label{fig:threeMethods}
\end{figure*}

In this article, we revisit the connection between gradients of local density and patterns of disclinations in ground states of 2D conformal crystals.  In particular, we aim to understand the role of two approximations underlying the proposed connection between disclination distributions of preferred density gradients in physical systems forming these structures.  This sequence of approximations show schematically in Fig. ~\ref{fig:threeMethods}.  The first is coarse-graining over discrete particle in the lattice-like order, replacing these with smoothly varying fields that model long-wavelength gradients in the local densities and bond-orientational order in the structure.  Such an approximation is standard to continuum theory descriptions of a variety of geometrically frustrated systems~\cite{nelson_defects_2002, Grason2016}, including liquid-crystalline and crystalline systems, and the accuracy of this approximation generically improves as the global dimensions far exceed the inter-particle spacing (i.e., the thermodynamic limit).   In this continuum-field description, topological defects  (e.g. disclinations and dislocations) correspond to order parameter configurations that are continuous except at singular point-like regions where it cannot single-valued.  In this way, these topological defects play the role of discrete ``quasi-particles'' to control underlying the long-wavelength order parameter distortions.  In this article, we call physical theory that are based on the {\it discrete} degrees of freedom of topological defects, the {\it Taylor} description, in analogy to the theory of solids based on stresses generated by individual and discrete dislocations~\cite{Taylor1934, Hirth1985}.  A second approximation, which we call the {\it Nye} description, assumes that defects are sufficiently numerous and reconfigurable that the they may be considered as continuous field itself~\cite{nye_geometrical_1953, kroner_leshouches_1981}.  The canonical example of this by Nye's definition of the ``geometrically necessary''  dislocation density needed to neutralize stress in a plastically deformed crystal, which predicts the local density and orientation of dislocations in terms of the row curvatures.  

In this language, the perfect neutralization of frustration by defects~\cite{Irvine2010, azadi_emergent_2014, Azadi2016} relies on the Nye description, assuming that the principle defects, disclinations, are well described by a continuous distributions that smear out their topological charge over large fractions of the crystal domain, as we outline in more detail below.  While intuitive and at least qualitatively correct for capturing certain features of defect pattern, there is reason to question about the accuracy of Nye picture when applied to multi-disclination ground states of conformal crystals.  For one, while dislocations are characterized by a microscopic size, the Burgers vector, which can be taken to be arbitrarily small compare the system size, the effective charge $Q$ per disclination defect, associated with the deficit or excess wedge angle in the Volterra construction $2 \pi Q/6$, cannot be arbitrarily small.  Hence, conformal crystal ground states do not in general possess a thermodynamically large total disclination charge even when the lattice spacing becomes vanishingly small compared to the system size.  As a consequence of this, conformal crystal ground states may be far more limited in terms of local density gradients that can be achieved by discrete disclination arrangements (i.e. in the Taylor description) than is implicitly assumed when considering them as a continuous distribution (i.e. the Nye description) that would perfectly neutralize the frustration imposed by non-uniform density.

In this article, we employ a 2D model of long-range repulsive particles subject to power-law confining potentials to investigate the optimal patterns of disclinations that form in their ground states based on the conformal crystal description.  Specifically, we compare the patterns of ground states predicted by three methods of calculation:  (i) numerical simulation of ground states (i.e. discrete particle packings with discretely distributed defects); (ii) a model of conformal crystals with discrete numbers of disclinations (i.e. continuum distribution of density with discretely distributed defects); and (iii) ``frustration-neutralizing'' predictions based on energetically optimal local particle density (i.e. continuum density and defect distributions).  These three class of models, which we compare in this study, are depicted schematically in Fig.~\ref{fig:threeMethods}.

Our focus on a model with screened 2D repulsions is motivated, on one hand, by the structure of confined assemblies of repulsive, line-like elements.  Examples of such systems include, vortex arrays in confined type-II superconductors~\cite{Geim1997, Baelus2004}, or confined arrays of stiff polyelectrolytes in an aqueous medium~\cite{Yao2013, podgornik_molecular_1990} (i.e. in the Debye-H\"uckel limit), in which mobile counterions screened long-electrostatic forces at long range.  More broadly, the choice of this class particle interactions also provides a convenient parameterization to explore the role of variable finite-range of repulsion on the emergent structure of the ground state.  For this model, we solve exactly for the energetically-optimal density patterns as a function of the screening length as well as the power law of the confining potential.  The optimal patterns of local particle density exhibit a dependence on repulsive range:  for long-range repulsions, local density increases with radial distance from the potential minimum; when repulsions become short-ranged, local density decreases with radial distance.  Consistent with the generic trends of the frustration-neutralizing picture~\cite{yao_topological_2013, Soni2018}, simulations and model predictions show that these two classes of particle density gradients are accompanied by respective excess in $Q<0$ (radially-increasing density) or $Q>0$ (radially-decreasing density) disclination defects in their predicted ground state structure.  A more detailed analysis of the total disclination charge as function of screening length shows a breakdown in the predictions of the frustration-neutralizing picture, which specifically dramatically overestimates growth of the number of $Q=+1$ (five-fold) disclinations with decreasing screening length.  

We then study a discrete-disclination model of conformal ground states, in which patterns of local density are fully determined by number, charge and spatial arrangement of multiple point-like disclinations.  This model predicts a drastically different pattern of optimal disclinations depending on the long- vs. short-range nature of repulsions and corresponding negative vs. positive charge of favorable defects in the packing:  highly-concentrated arrangements of $Q=-1$ defects populate the centers of long-range repulsive clusters; while split arrays of $Q=+1$ decorate edges of clusters when interactions become short-ranged.  We show that this discrete-defect theory correctly captures the saturation of total defect charge in the limit of vanishing screening length observed in the simulated ground states.  Based on this comparison, we argue that a naive application of frustration-neutralizing approximation (i.e. continuously distributed defects) fails to accurately predict the number positive defects that form in the limit of short-range interactions for two reasons.  The approximation fails to capture the position-dependence of dislination energetics, and further, it does not account for anharmonic dependence of the defect energetics on disclination charge (i.e. self-energies remain finite for $Q \ll -1$ but diverge $Q \gg +1$).

The remainder of this article is organized as follows.  We begin in Sec.~\ref{sec: conformal} with an overview of the conformal crystal model, and in particular, the condition that relates disclination charges to particle density gradients.  We  review the assumptions of the frustration-neutralizing prediction for the continuum distribution of disclinations based on the energetically optimal particle density.  In Sec. ~\ref{sec: models} we introduce the our 2D model assembly, specifically in the context of parallel and line-like columnar elements in confined in (cylindrical) power-law potentials interacting through (2D) screened repulsions.  We describe the exact solutions for the energetically optimal particle densities assuming continuously distributed defects in Sec.~\ref{sec: continuum}, followed by our approach to the discrete disclination model of ground-state energetics in Sec.~\ref{sec: discretedefect} and numerical simulation methods for discrete-particle ground states in Sec.~\ref{sec: simulation}.  In Sec.~\ref{sec: results}, we describe the comparison of fluid density predictions to simulated ground state structure for both quadratic and quartic potentials, specifically contrasting the emergent defect patterns for long-range and short-range repulsion, characterized respectively by excess negative and positive disclinations.  We then contrast these results with predictions of the discrete defect model in  Sec.~\ref{sec: discrete}, and focus on the role of localization and spreading of discrete disclinations in optimal structures.  In Sec.~\ref{sec: discussion} we discuss the origins for the contrasting behaviors in terms of an asymmetric dependence of defect energetics on the sign of the disclination charge in Sec.~\ref{sec: discrete}.

\section{\label{sec: conformal}Elements of the conformal crystal model}

In this section, we overview the theoretical elements of the conformal crystal model and its connection to ground states of confined, long-range repulsive particles.  The notion of a conformal crystal was introduced by Rothen, Pieranski, Rivier and Joyet~\cite{Rothen_1993, Rothen_1996}, and its has been described and applied to numerous physical scenarios~\cite{mughal_topological_2007,yao_topological_2013, Soni2018, Menezes2017, Straube2019}.  Here, we summarize the key assumptions of the model and the mathematical connection between local density, lattice curvature and disclination defects.

This model is most conveniently described in terms of complex functions that describe mappings of 2D plane and well known mathematical properties of analytic functions (see e.g. \cite{Needham2002}).  A conformal crystal is defined as a {\it conformal mapping} of a regular 2D lattice (the reference state) to distorted configuration on the plane (the conformal crystal)~\cite{Rothen_1993}. The distortions of the conformal map correspond to locally-isotropic swelling of the neighbor spacing the lattice, and further more, the map is angle preserving, such that the local bond angles in the conformal crystal are preserved from the uniform reference state.  Specifically, we consider in an initially equi-triangular lattice for the reference state, in which case, the conformal crystal corresponds to a state where six neighbors are (nearly) equally-distributed around (almost) every point.  

The motivation for the conformal crystal model derives largely from observations from experiment or simulation suggest that ground states adopt locally isotropic arrangements, in which neighbors are everywhere symmetrically arranged around every particle, but with long-range gradients in neighbor spacing.  Physically, this condition most likely arises from the near-field effects of repulsive interactions with neighbors.  In most cases repulsions grow large (if not, diverge) with vanishing separation (e.g. Coulomb or dipolar), so that to a first approximation forces on particles are dominated by the nearest neighbors, and force balance on each particle generically satisfied by symmetric (i.e. $n$-fold) arrangements of equally spaced neighbors.  Of course, this argument ignores the effects of longer range gradients in the pressure distribution arise from an external potential and the long-range effects of interaction.  And hence, one should expect the locally-isotropic condition to strictly only apply in the continuum limit where the neighbor spacing becomes arbitrarily small compared to the scale of gradients in the pressure or density.

The structure of conformal crystals and their defects is defined by conformal map that takes a point in the uniform reference lattice ${\bf r}_0= (x_0,y_0)$ to its location in the conformal crystal ${\bf r}=(x,y$).  Using the notation of the complex plane, these points are $z_0= x_0 + i y_0$ and $z = x + iy$, and well-known properties of analytic functions, it is straightforward to show that if $z(z_0)$ analytic function, that the map is conformal, or ``locally-isotropic'' in the sense introduced above.  As described in the Appendix \ref{app:proof}, this can be captured by the deformation gradient, that describes the map from length element $dz_0$ in the uniform lattice to the conformally-deformed one $dz$,
\begin{equation}
dz = \Omega (z) e^{i \theta (z) } ~dz_0 
\end{equation}
where $ \Omega (z)$ and $\theta (z)$ describe, respectively, the local swelling and rotation angle of patch at $z$.  If the density of the reference lattice is $\rho_0$, this density in the conformal lattice is
\begin{equation}
\rho(z) = \Omega^{-2} (z)  \rho_0 .
\end{equation}
Because $\Omega(z)$ is also an analytic function, it can can be shown that (see Appendix \ref{app:proof}) that gradients in angle and density are not independent.  Expressing this condition in terms of functions of Cartesian coordinate, ${\bf r} = x ~\hat{x} +y ~\hat{y}$, gives
\begin{equation}
\label{eq: benddensity}
\grad_\perp \theta( \rv) = \frac{1}{2} \hat{z} \times \grad_\perp \ln \rho (\rv) ,
\end{equation}
where $\grad_\perp = \hat{x} \partial_x + \hat{y} \partial_y$ is the in-plane gradient and $\hat{z} = \hat{x} \times \hat{y} $ is the plane normal.  Gradients in the $\theta(\rv)$ imply the spatial variation of bond directions, and hence,  curvature of the rows along the direction of $\grad_\perp \theta( \rv) $.  Hence, eq. (\ref{eq: benddensity}) implies that bending of the rows of the conformal crystals is accompanied by proportional gradients in the logarithm of the local density normal to the rows.

To relate this condition to areal distribution of disclinations, $s(\rv)$, we consider a changes $\theta( \rv)$ around closed loops, or Burgers circuits~\cite{nelson_defects_2002}, enclosing a region ${\cal D}$ of the conformal lattice.   In order that the conformal distortion be consistent with a well-defined (six-fold) bond order parameter away from disclination cores, changes in the  $\theta(\rv)$ around any closed loop must occur in integer multiples of $2 \pi/6$.  That is,
\begin{equation}
\label{eq: stokes}
\oint_{\partial {\cal D}} d {\bf s}  \cdot \grad_\perp \theta( \rv) = \int_{{\cal D}} d^2\rv ~ \Big(\frac{1}{2} \grad_\perp^2 \ln \rho \Big) = \frac{2 \pi}{6} Q_{{\cal D}}
\end{equation}
where we have used eq. (\ref{eq: benddensity}) Stokes law convert contour integer into an integral over the area of ${\cal D}$.  Here, $Q_{{\cal D}}$ is an integer reflecting the total charge of disclinations enclosed in ${\cal D}$, which does not change with any continuous change of $\partial {\cal D}$ that the encloses the same set of defects.  Accordingly, the area integrand of eq. (\ref{eq: stokes}) must define the areal density of disclinations from eq. (\ref{eq: stokes}),
\begin{equation}
\label{eq: conformal}
-\frac{1}{2} \grad_\perp^2 \ln \rho = s (\rv)
\end{equation}
where
\begin{equation}
\label{eq: disclinations}
s (\rv) = \sum_\alpha s_\alpha \delta^{(2)} (\rv -\rv_\alpha) ,
\end{equation}
is the sum of point-wise disclinations at positions $\rv_\alpha$ and with charges $s_\alpha = 2 \pi Q_\alpha/6$, where $Q_\alpha = 0, \pm 1, \pm 2, \ldots$ is required for single-valued six-fold bond order away from the disclination cores.  In this way, eqs. (\ref{eq: conformal}) and (\ref{eq: disclinations}) show that disclinations act as ``monopole sources'' for density gradients in conformal crystals, with $(\ln \rho) /2$ playing the role of a potential.  These equations serve as constraints on the spatial patterns of density variation that are possible in conformal crystals.

One approach to understand the structure of conformal crystal states of confine repulsive particles, outlined first by Mughal and Moore~\cite{mughal_topological_2007}, is to consider a continuum variational approach to the optimal patterns of density.  In this approach, we have a simplified form of the energy functional in terms of particle density,
\begin{multline}
\label{eq: continuum}
E_{cont}[ \rho(\rv)] = \frac{1}{2} \int d^2\rv \int d^2\rv' ~\rho(\rv) V_{int} (|\rv-\rv'|) \rho(\rv') \\ + \int \int d^2\rv ~ U (\rv) \rho(\rv)
\end{multline}
where $V_{int}(r)$ is the pair potential, while $U (\rv)$ describes a spatially-confining potential field.  Ignoring any constraints on the density imposed by local-correlations in the packing, this energy is readily optimized subject to the constraint of fixed particle number,
\begin{equation}
\label{eq: number}
N = \int d^2 \rv ~ \rho (\rv) ,
\end{equation}
yielding an self-consistency condition in terms of a spatially constant chemical potential 
\begin{equation}
\label{eq: mu}
\mu = \int d^2\rv' ~V_{int} (|\rv-\rv'|) \rho_{f}(\rv') + U(\rv) .
\end{equation}
Here $\mu$ is Lagrange multiplier that is chosen to set $N$, and $\rho_{f}(\rv)$ refers to the density pattern that minimizes $E_{cont}$, which we call the optimal {\it fluid} density, since no constraints are imposed on gradients of $\rho_{f}(\rv)$, i.e. through the combined conditions of eqs. (\ref{eq: conformal}) and (\ref{eq: disclinations}).

Notwithstanding the fact that the continuum formulation of eq. (\ref{eq: continuum}) does not enforce the local correlations of the conformal crystals (i.e. locally isotropic packing) nor discreteness of particles themselves, one approach has been to use this optimal fluid density in combination with the relationship between $\rho(\rv)$ and defects in conformal crystals in eq. (\ref{eq: conformal}) to predict the expected distributions of disclinations in their ground states~\cite{mughal_topological_2007}.  This gives a {\it fluid} disclination density 
\begin{equation}
\label{eq: fluiddefect}
s_f(\rv) = - \frac{1}{2} \grad_\perp^2 \ln \rho_f .
\end{equation}
The fluid dislocation distribution is equivalent to the frustration distribution of disclinations proposed in refs. \cite{yao_topological_2013, mughal_topological_2007} , and one can view this that Nye picture of disclinations, which assumes that defects are sufficiently numerous and mobile that effectively their arrangements may give rise to whatever pattern of fluid density that optimizes the $E_{cont}$.  This prediction (or its equivalent) has been tested against several models of confined and long-range repulsive particles, and to some extent, this relation works well to predict at least the sign and spatial distribution of disclinations that are observed, in numerically simulated ground states.

In the following, we explore the validity of this fluid defect density model for particles repelling through screened 2D Coulomb repulsions and test the underlying assumption that optimal fluid density patterns can, in general, be realized in (locally-isotropic) conformal crystals.

\section{\label{sec: models}Confined ground states of screened, 2D Coulomb repulsions  }

We investigate a general class of for 2D ground states $N$ repulsive particles confined by an external potential, $U(\rv)$
\begin{eqnarray}
\label{eq: totale}
E=\sum\limits_{i<j}^N V_{int}(|\mathbf{r_i}-\mathbf{r_j}|) + \sum\limits_{i}^{N}U(\mathbf{r_i})  \label{eq:discreteEnergy} ,
\end{eqnarray}
where ${\bf r}_i$ is the position of the $i$th particle.  Here, we consider a repulsions corresponding to screened Coulomb interactions,
\begin{equation}
\label{eq: Vint}
V_{int}(r) = v_0 K_0(\kappa r)
\end{equation}
where $K_0(\cdot)$ is the modified Bessel function of the second kind, $v_0$ parameterizes the strength of repulsions (e.g. linear charge density of parallel filaments) and $\kappa^{-1}$ parameterizes the {\it screening length} of repulsions (i.e. crossover from logarithmic, $V_{int}(\kappa r \ll 1 )\simeq -v_0 \ln (\kappa r)$, to exponential $V_{int}(\kappa r \gg 1)\simeq v_0 e^{-\kappa r}$ decay).  Here we consider axisymmetric confining potentials, $U(\rv) = U(|\rv|)$, where potential increases with distance from the origin $\rv =0$, e.g. the center of a confining field.  In what follows study two examples of confining potentials of power law form
\begin{equation}
\label{eq: powerlaw}
U(r)=u_0 r^n
\end{equation}
where $n>0$ and $u_0$ parameterizes the confinement strength.  In addition to the screening length, $\kappa^{-1}$, this ratio between the strength of interactions to confining potential defines the characteristic length scale,
\begin{equation}
\ell \equiv  (v_0/u_0)^{1/n} .
\end{equation}
As described below, it is convenient to parameterize units of length in numerical simulation in units $\ell$. 

We employ and compare three approaches to the optimal defect structure in ground states of eq. (\ref{eq: totale}): fluid density (continuum defect) model; discrete defects in conformal crystals; and numerical simulations of discrete particles.

\subsection{ Fluid density model for disclination density }
\label{sec: continuum}

Here we extend the variational fluid density approach Mughal and Moore~\cite{mughal_topological_2007} to the case of screened Coulomb repulsions in axisymmetric confining potentials.  As detailed in the Appendix \ref{app:K0Cal}, the optimal fluid density $\rho_f(r)$ solved for arbitrary screening length and axisymmetric confining potential, where $r$ is the radial distance from the center of the potential (i.e. where $\nabla_\perp U =0$).  This method relies on recasting the self-consistent chemical potential in eq. (\ref{eq: mu}) as
\begin{equation}
\label{eq: mu2}
\mu = U(r) + \phi (r), \ \ \ \ { \rm for \ } r \leq R
\end{equation}
where $R$ is finite size of the confined density (i.e. $\rho_f (r>R) =0$) and $\phi(r)$ is the potential generated by repulsions
\begin{equation}
\label{eq: phi}
\phi (\rv) = \int d^2 \rv' ~ V_{int} (|\rv-\rv'|) \rho_f (\rv') .
\end{equation}
Here, it is convenient to parameterize the solutions in terms of optimal radius of clusters $R$, which effectively controls the total particle number $N$.  Using the fact that $V_{int}(\rv)$ is the Green's function for 2D screened electrostatics --- i.e. $(\nabla_\perp^2 + \kappa^2) V_{int}(\rv) = 2 \pi v_0 \delta^{(2)} (\rv)$ -- it is straightforward to show that eqs. (\ref{eq: mu2}) and (\ref{eq: phi}) are solved by
\begin{equation}
\rho_f(r) = \frac{1}{2 \pi v_0} \Big\{ \kappa^2 \big[ \mu - U(r) \big] + U''(r)+r^{-1}  U'(r) \Big\} 
\label{eq:rhoGeneralFinal}
\end{equation}
where $U'(r) = \partial_r U(r)$.  The chemical potential is linked self-consistently to the potential field $\phi(r)$  via eq. (\ref{eq: phi}) yielding, 
\begin{equation}
\mu(R) = \frac{2 R I_0(\kappa R) K_0(\kappa R) U'(R)}{1 - \kappa R[I_1(\kappa R)K_0(\kappa R)-I_0(\kappa R)K_1(\kappa R)]} + U(R) \label{eq: muR} ,
\end{equation}
where $I_n( \cdot)$ is the modified Bessel function of the first kind.  The relationship between particle number and cluster size follows directly from the integral $N(R) = 2 \pi \int_0^R dr~r \rho_f(r)$.

Applying these results to the case of power-law potentials, $U(r) = u_0 r^n$, we find the optimal fluid density patterns
\begin{equation}
\label{eq: rhof}
\rho_f(r) = \rho_0 \big[ f_n(\kappa R) + n^2 (\kappa r)^{n-2}-(\kappa r)^n \big] ,
\end{equation}
where
\begin{equation}
\rho_0 = \frac{ u_0}{2 \pi v_0} \kappa^{2 - n} = \frac{\kappa^2}{2 \pi (\kappa \ell)^n },
\end{equation}
and
\begin{equation}
\label{eq: fR}
f_n(x) \equiv x^n\Big[1+\frac{2 n I_0(x) K_0(x)}{1-x[I_1(x) K_0(x)-I_0(x)K_1(x)]}\Big] . 
\end{equation}
The total particle number (as a function of optimal cluster radius) takes the form,
\begin{equation}
\label{eq: NR}
N= \rho_0 \kappa^{-2} g_n(\kappa R) ,
\end{equation}
where
\begin{equation}
\label{eq: gR}
g_n(x) \equiv \frac{1}{2}x^2 f_n(x) +\Big(n-\frac{x}{n+2}\Big) x^n .
\end{equation}
Notably this shows that the reduced size $\kappa R$ of the cluster is only function of a single parameter, the {\it scaled particle number} 
\begin{equation}
\label{eq: barN}
\bar{N} \equiv N (\kappa \ell)^n=  N (\kappa^n v_0/u_0) ,
\end{equation}
via the equation of state
\begin{equation}
\label{eq: kRvsN}
\bar{N} = g_n(\kappa R) .
\end{equation}

Before moving onto to the prediction of fluid defect density, we briefly comment on general dependence of $\rho_f(r)$ on the range of interactions and the natural of the confining potential (i.e. the power law $n$).  From eq. (\ref{eq: rhof}) the optimal density includes two spatially varying terms.  The first of these grows as $n^2 r^{n-2}$ is non-decreasing for $n \geq 2$ (i.e. quadratic or steeper), while the second decreases as $-\kappa^2 r^n$.  The relative dominance of these two terms is determined by radius:  for $r \ll n \kappa^{-1}$ the density is non-decreasing, while for $r \gg n \kappa^{-1}$ the density decreases with radius.  

To understand the physical origins of these competing terms, it is instructive to consider the limiting cases of infinitely long- and short-range interactions.  In the former case, where $\kappa \to 0$, repulsive forces reduces to 2D electrostatics.  Assuming a power-law density profile $\rho_f(r) \propto r^\alpha$, Gauss's law gives an outward force from interactions $F_{int} \propto r^{\alpha+1}$, which is balanced by the inward force of potential, $F_{pot} = - \partial_r U \propto-n r^{n-1}$.  Force balance then requires $\alpha = n -2$ and
\begin{equation}
\rho_f(r) \propto r^{n-2} \ \ \ \ {\rm for} \ \kappa \to 0 ,
\end{equation}
illustrating the general tendency for long-range repulsions to spread density out to large radii (i.e. $\rho_f(r)$ is {\it convex}).  Alternately, in the limit with $\kappa \to \infty$, interactions become localized such that potential becomes proportional to the local density, $\phi(\rv) \propto v_0 \kappa^{-2} \rho_f (\rv)$, indicative of the local compressibility of the fluid.  From eq. (\ref{eq: phi}), we then have
\begin{equation}
\rho_f \propto \kappa^2 \big[ \mu -  U(r) \big] \propto C_0 - r^n  \ \ \ \ {\rm for} \ \kappa \to \infty ,
\end{equation}
which illustrates the countervailing tendency of the confining potential to {\it concentrate} particle density towards its minimum when interactions become short ranged. Hence, as we observe below, as the interactions vary from long-ranged ($\kappa R \to 0$) to short-ranged ($\kappa R \to \infty$), the optimal fluid density varies from non-convex to convex, respectively, corresponding to transitions in the charge, number and spatial pattern expected from disclinations.

From eq. (\ref{eq: fluiddefect}) the fluid disclination profile is
\begin{equation}
s_f(x=\kappa r) =\kappa^2 n^2 x^n \frac{4 x^n+f_n(\kappa R) \big[x^2-(n-2)^2\big]}{2\big[x^2 f_n(\kappa R)+(n^2-x^2) x^n\big]^2 } .
\end{equation}
Averaging this distribution over the cluster yields the total fluid disclination number,
\begin{eqnarray}
\label{eq: Qf}
Q_f &=& 6 \int_0^R dr ~r s_f( \kappa r) = -3 R \frac{ \rho_f'(R)}{\rho_f(R)} \nonumber \\ 
&=& -3 n \frac{(n-2) n (\kappa R)^{n-2} -  (\kappa R)^n   }{f_n(\kappa R) + n^2 (\kappa R)^{n-2}-(\kappa R)^n } ,
\end{eqnarray}
where we have used Gauss's law and eq. (\ref{eq: fluiddefect}).

Returning again to the limiting case of long-range interactions, we can show that predicted fluid disclination density becomes,
\begin{equation}
\label{eq: skappa0}
s_f(\rv) = -\pi (n-2) \delta^{(2)}(\rv) , \ \ \ \ \kappa R \to 0 ,
\end{equation}
with non-positive disclinations (for $n >2$) concentrated at the core of the cluster.  In the opposite limit of short-range interactions, fluid disclination density becomes 
\begin{equation}
s_f(r) = \frac{n^2 r^{n-2} R^n}{2 (R^n-r^n)^2} , \ \ \ \ \kappa R \to \infty ,
\end{equation}
which is positive and concentrated at the rim of the cluster $r \to R$.  Considering the total disclination number in the limits of long-range interactions, $Q_f$ approaches a minimal value $Q_f( \kappa R \to 0) \to -3 (n-2)$, while in the opposite limit of short range interactions, the total fluid disclination prediction is positive and grows unbounded, as $Q_f( \kappa R \to \infty) \to 3 \kappa R$.  We return these predictions for ground-state defect patterns below. 

\subsection{ Discrete disclinations in conformal crystals }
\label{sec: discretedefect}

The fluid density model of the previous section considers that minimal energy pattern of particle density independent of the constraints imposed by local correlations in the conformal crystal.  Here, we consider a second approach to model the ground state energetics that explicitly constructs the particle density patterns enforcing geometric connection between density gradients and density of spatially-localized and quantized topological charge disclinations.   Specifically, we consider density patterns $\rho_d (\rv)$ that are parametrized by the distributions of disclinations $s (\rv) = \sum_\alpha s_\alpha \delta^{(2)} (\rv -\rv_\alpha) $, and then optimize $E_{cont}$ with respect to number, charge and positions of defects in the cluster.  This {\it mesoscopic} defect theory accounts for the fact that multi-disclination conformal crystals may or may not actually realize the optimal fluid density patterns $\rho_f(\rv)$.

To construct this theory we rewrite eq. (\ref{eq: conformal}) in the form of a Poisson equation
\begin{equation}
\label{eq: psi}
\grad_\perp^2 \psi(\rv) = s (\rv),
\end{equation}
where $\psi(\rv)$ is a potential related to the density
\begin{equation}
\rho_d(\rv) = e^{-2 \psi (\rv) } .
\end{equation}
For our calculations we will assume that the boundary of the cluster maintains a constant circular radius of $r=R$, which we find to be in good agreement with simulated clusters as described below.  Boundary conditions are required to solve eq. (\ref{eq: psi}) for the density given a set of disclinations.   Here, we opt for Dirichlet boundary conditions, which require the density (and $\psi(\rv)$) to be constant on boundary.  This condition is consistent with the axisymmetry of the optimal density fluid density patterns at the outer edge.  Beyond this, by Chebyshev's principle~\cite{bermejo_minimum_2005}, Direchlet conditions minimize the variation of density in simply connected (i.e. defect-free) domains, and conjectured to do so in multiply connected domains.  As the variation of density in the defect cores is shown to be energetically costly, we conjecture that such a condition should be favored by minimization of $E_{cont}$.  

With these boundary conditions, multi-disclination solutions can be constructed by superposition.  The Green's function for Poisson's equation with Direchlet boundary conditions on the 2D disc and a monopole source at $\rv_0$ has the form
\begin{equation}
G(\rv,\rv_0) = \frac{1}{4 \pi} \ln \Big[\frac{R^2+r^2 r_0^2/R^2-2r r_0 \cos (\theta - \theta_0) }{r^2+r_0^2-2r r_0 \cos (\theta - \theta_0)} \Big] ,
\end{equation}
where $(r,\theta)$ are polar coordinates.  From this, the general solution for $\phi(\rv)$ takes the form
\begin{equation}
\psi(\rv) \equiv \psi(R) - \delta \psi(R) ,
\end{equation}
where
\begin{multline}
\label{eq: deltapsi}
\delta \psi (\rv) = \int d^2 \rv' ~ G(\rv,\rv') s (\rv ') = \sum_{\alpha} G(\rv,\rv_\alpha) s_\alpha
\end{multline}
and $\psi(R)$ is (constant) value at $r=R$ since $ G(\rv,\rv')$ for $r\to R$.  From this we have the density of the conformal crystal in terms of defect positions and charges
\begin{equation}
\rho_d(\rv) = \rho_R \exp\big[2 \sum_\alpha  s_\alpha G(\rv,\rv_\alpha) \big] ,
\end{equation}
where $\rho_R = e^{-2 \psi(R)}$ is the density at the cluster boundary, which is set by normalization,
\begin{equation}
\rho_R = \frac{N}{\int_0^R dr ~r \int_0^{2 \pi} d\theta ~  e^{\big[2 \sum_\alpha   s_\alpha G(\rv,\rv_\alpha) \big]}  } .
\end{equation}
Given these solutions, the energy of interactions and confinement is computed by inserting $\rho_d(\rv)$ into eq. (\ref{eq: continuum}), and minimizing $E_{cont}$ over $R$.  

For the general non-axisymmetric defect configurations, determining $\rho_d$, computing its energetics, and minimizing of cluster size requires numerical integration over the disc area.  It is instructive to consider the axisymmetric solution of $\rho_{axi}(\rv)$ when defects are confined to origin, also presented in ref. \cite{mughal_topological_2007}.  Here we consider a defect of charge $s_0 = \tfrac{\pi }{3} Q_0$ at $r=0$.  The axisymmetric potential is $\psi(r) = Q_0\tfrac{\pi}{6} \ln(r/R) + \psi(R)$, corresponding to density
\begin{equation}
\label{eq: axidisc}
\rho_{axi} (\rv) = N \frac{ (1-Q_0 \pi/6)}{\pi R^2} \Big( \frac{r}{R}\Big)^{-Q_0/3} ,
\end{equation}
where $e^{-2 \psi(R)} =  N(1-Q_0/6)/(\pi R^2)$ is set by normalization.  This power-law form for the density profile illustrates the geometric distinction between positive and negative disclinations.  For $Q_0<0$, the density is concave, decreases with radius and is finite everywhere on the disc.  In contrast, for positive disclinations, the density is convex, decreases with radius and diverges as the defect core.  We note further that this central monopole solution agrees exactly with the fluid disclination density predicted for the limit of long-range interactions, in eq. (\ref{eq: skappa0}), which for smooth confining fields ($n\geq2$) corresponds non-positive topological charge.  We return to the consequences of this asymmetric dependence on sign of topological charge below.

\begin{figure*}[t]
    \centering
    \includegraphics[width=0.98\textwidth]{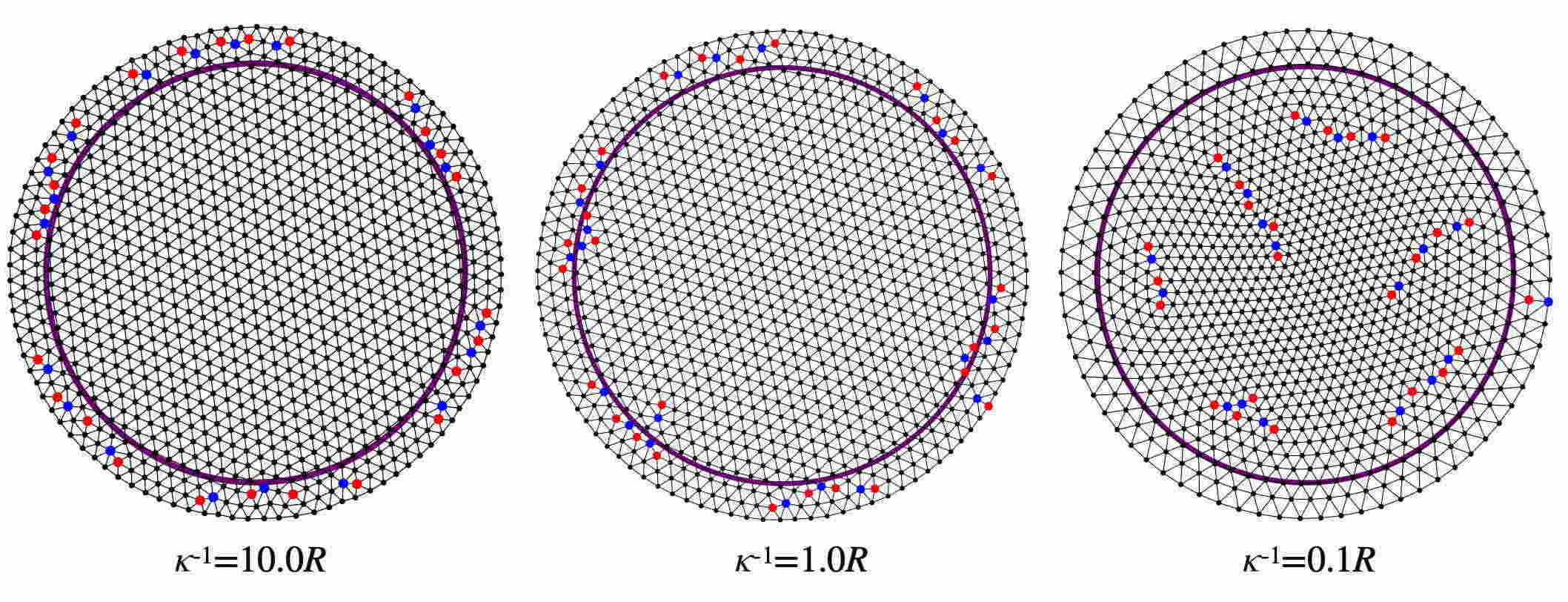}
    \caption{Simulated ground state configurations for harmonically confined clusters of $N = 500$ with variable screening length $\kappa^{-1} =  10.0R,  1.0R,  0.1R$.  Delaunay triangulation shows the neighbor network, with site coordination indicated by black (6), red (5) and blue (7).  The purple circle denotes what is used that define the ``interior" region for bulk disclination charge count, $85\%$ of the radial distance to the outer boundary.}
    \label{fig:r2snapshot}
\end{figure*}

\subsection{ Numerical simulation of ground states }
\label{sec: simulation}

To test and compare the validity of the two theoretical models for defect ground states in conformal crystals, we carry out numerical simulations of confined ground state clusters of the 2D screened, Coulomb particle model. We consider clusters of a fixed number of particles, $N = 100, 200, 500, 1000, 2000$ or $5000$ , confined to either quartic and parabolic potential.  

The objective of the simulations is to survey optimal defect configurations spanning from long-range ($\kappa R \ll1$) to short-range ($\kappa R \gg 1$) interactions.  As the equilibrium radius $R$ is not fixed in the simulations, we instead define our simulation parameters based on the predicted dependence of cluster size on particle and interactions.  Specifically, for a given number of particles, $N$, we chose normalized screening lengths $\kappa \ell$, such that we could target fluid density cluster sizes over the range $\kappa R = 0.1$ to 30.  

For each target $\kappa R$ and $N$, we generated 600 random initial configurations and find the lowest energy configuration via multidimensional minimization, where initial positions are randomly placed within area that is 10 times the dimension of the target equilibrium cluster size (based on fluid density model predictions).  After testing three minimization algorithms -- conjugate gradient~\cite{plimpton_fast_1995}, FIRE\cite{bitzek_structural_2006} and quickmin\cite{sheppard_optimization_2008} -- we find that FIRE leads to the lowest energy and force residuals, and therefore, implement this for our model with LAMMPS~\cite{lammps}.  The stopping criteria is that the total force residual at the whole system is less than $10^{-5} ~ v_0/\ell$ for $N=100, 200, 500$ and $10^{-4}~ v_0/\ell$ for $N=1000, 2000, 5000$.  For each resulting structure the equilibrium radius $R$ of the cluster is determined by the radial position of the outermost filament. 

We select the resultant configuration with the lowest energy for each parameter, analyze its packing via Delaunay triangulation~\cite{barber_quickhull_1996} and identify the outer edge of cluster, from which the mean radius $R$ of the cluster can be measures.  Disclinations in the structure are identified as points that deviate from 6-fold bond coordination.  For particles in the interior of the cluster, the charge is $Q=6-z$, where $z$ is the local coordination of the vertex in the bond network.  For sites on the edge, we take the definition of edge disclinations as $Q=4-z$, which is known to be consistent with fixed Euler characteristic of the disc, i.e. the total disclination charge, interior plus boundary, remains +6~\cite{Bowick2009}.

We also extract a measure of the local density in the simulated packing via the Voronoi tesselation, which is dual to the Delaunay triangulation.  If particle $i$ at $\rv_i$ has a Voronoi polygon of area $A_i$, we define the local density as,
\begin{equation}
\rho_{local}(\rv_i) = A_i^{-1};
\end{equation}
Analyzing the distribution of local density over the cluster provides a direct comparison to predicted particle density patterns in the fluid density and discrete defect theories described above.

\section{\label{sec: results}Discrete particle ground states vs. fluid density model}

In this section we present and compare results of ground-state simulations and the fluid density model (i.e. continuous disclinations as in the Nye picture) in terms of the gradient patterns of particle density as well as defect distributions.  We consider two particular cases of power law confining potentials, harmonic ($n=2$) and anharmonic, or quartic, ($n=4$), which differ in terms of the relative tendency to concentrate particles at the potential minimum.  For both confinement types we consider the variation of ground state structure with interaction range, from $\kappa R \ll 1$ to $\kappa R \gg 1$, which we show has the affect reducing the tendency of repulsions to spread the distribution.  In the subsequent section, we discuss the comparison to predictions of the discrete disclination model introduced in Sec. ~\ref{sec: discretedefect}.

\subsection{Harmonic confinement (quadratic potential)}

We begin with the case $n=2$ confinement.  Using the results for the fluid density model in Sec. \ref{sec: continuum} above, the predicted fluid density follows an concave, inverse parabolic form 
\begin{equation}
\label{eq: rhofn2}
\rho_f(r)  = \rho_0\big[4 + f_2 (\kappa R) - (\kappa r)^2 \big]  \ \ \ \ {\rm for \ } n =2 ,
\end{equation}
where $f_2(x)$ is defined in eq. (\ref{eq: fR}).  Hence the density always predicted to decrease with $r$, but the rate of decrease is controlled by the screening.  The density tends toward uniform $\rho_f$ in the $\kappa R \to 0$ limit of long-range repulsions and in the opposite limit, $\kappa R \to \infty$, the density vanishes at the cluster edge, $\rho_f(r)  \simeq \rho_0\kappa^2 (R^2 -r^2)$.  Corresponding to this inverted parabolic density is the predicted defect density,
\begin{equation}
s_f (r) =2 \kappa^2 \frac{ \rho_0 \rho_f(0)}{\rho_f^2(r)} \ \ \ \ {\rm for \ } n =2 ,
\end{equation}
which implies a radially increasing density of positive (i.e. 5-fold) disclinations.  Averaging over the cluster area, we have to total predicted disclination charge,
\begin{equation}
Q_f =6  (\kappa R)^2 \frac{\rho_0}{\rho_f(R)} \ \ \ \ {\rm for \ } n =2 \label{eq:Qpara},
\end{equation}
As shown below in Fig. \ref{fig: parabolicsim}B, the predicted total fluid disclination charge vanishes for long-range interactions, consistent with a uniform density packing.  However, because fluid disclination density diverges at the cluster edge as $s_f (r \to R) \sim 1/(1 -r/R)^2$ in the limit of short-range interactions, the total charge is also predicted to be unbounded in this limit, $Q_f (\kappa R \to \infty) \to 3 \kappa R$.

In Fig. \ref{fig:r2snapshot} examples are shown from simulated ground states $N=500$ harmonically confined particles, ranging from long-range to short-range repulsions.  For long-range and intermediate-range cases, the interior of the cluster possesses nearly uniform, equitriangular order.  Because the cluster boundaries adopt a circular symmetry, the net 6 ``boundary disclinations'' of a disc-like domain, are retained within the one or two layers of the cluster boundary~\cite{Bowick2009}. For the case of short-range interactions $\kappa^{-1} = 0.1~R$, we observe visible gradients in the local density along the radial direction.  Simultaneously we observe that multiple strings of alternating 5-7 defects, sometimes known as  defect ``scars", are incorporated in the interior of the cluster.  Careful inspection shows that each of these structures each has an excess 5-fold defect, and possesses a net $Q=+1$ disclination charge.  

In Fig. \ref{fig: parabolicsim}A, we compare the equation of state relating equilibrium size of clusters to scaled particle number, eq. (\ref{eq: kRvsN}), from the fluid density model to the measured size of simulated clusters.  Simulation results largely confirm the prediction that reduced size $\kappa R$  of the clusters is only a function of scaled particle number $\bar{N}$ over the full range of screening length to cluster size ratios.  The fluid density model somewhat overpredicts cluster radius in the short-range limit (i.e. $\kappa R \gg 1$), for fairly small $N$ values, but as $N$ increases the simulated results tend to converge to the fluid density predictions. 

In Fig. \ref{fig: parabolicsim}B, we plot the defect charge $Q$ in the interior of the simulated clusters as function of $\kappa R$ and for a range of particle numbers from $N=100$ to $N=5000$.  Again, to distinguish from boundary defects within the first few layers of the cluster, we count interior disclinations as those within the inner 85\% of radial thickness of the cluster (see Appendix \ref{app:countQ} for detailed discussion).  We also plot the prediction of total defect count from the fluid density model, $Q_f$ for comparison.  For small clusters ($N =100, 200$), the boundary threshold introduces some ambiguities in distinguishing from between bulk and boundary of the cluster, due to the fact that the cluster possesses only a few layers of particles in the radial direction.  Notwithstanding this ambiguity,  larger cluster sizes tend to agree with the fluid defect prediction and are defect-free in their interiors in the long-range repulsive regime, $\kappa R \ll1$.  

\begin{figure}
    \centering
    \includegraphics[width=0.48\textwidth]{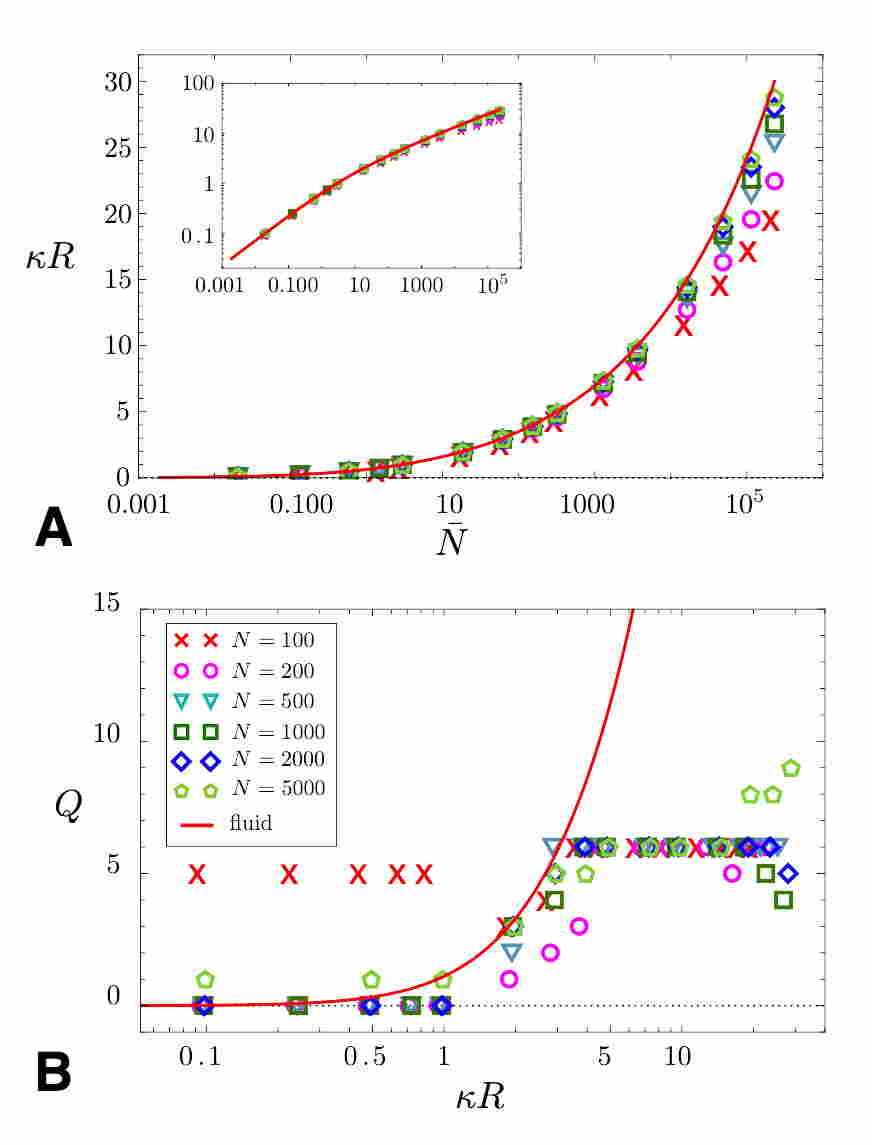}
    \caption{Comparison of size (A) and disclination charge (B) between fluid density model and simulated (discrete) particle ground states for harmonic confinement ($n=2$).  In (A), comparison of the equation of state in eq. (\ref{eq: kRvsN}) for scaled cluster size vs. scale particle in the fluid density model to the equilibrium cluster size of simulated clusters.  In (B), the predicted disclination charge in the fluid density model compared to interior defect count (within the 85\% of the radial distance from the center), both plotted as functions of the reduced cluster size $\kappa R$.  The numbers of particles in simulate clusters are indicated by the legend in (B).  }
    \label{fig: parabolicsim}
\end{figure} 

As the screening length decreases, the interior disclination charge $Q$ of simulated clusters increases, due to the excess of interior 5-fold defects, qualitatively consistent with fluid defect prediction.  However, while $Q_f$ grows arbitrarily large as $\kappa R \to \infty$, we observe that interior defect charges only increases to a {\it finite value} of $Q\simeq 6$.  For the exceptional case of the largest particle number, $N=5000$, we do observe up to $Q =9$ for the largest value of $\kappa R \simeq 30$, this falls far below the fluid defect prediction $Q_f $ which exceeds $+10^3$ for this interactions range.

\begin{figure}
    \centering
    \includegraphics[width=0.48\textwidth]{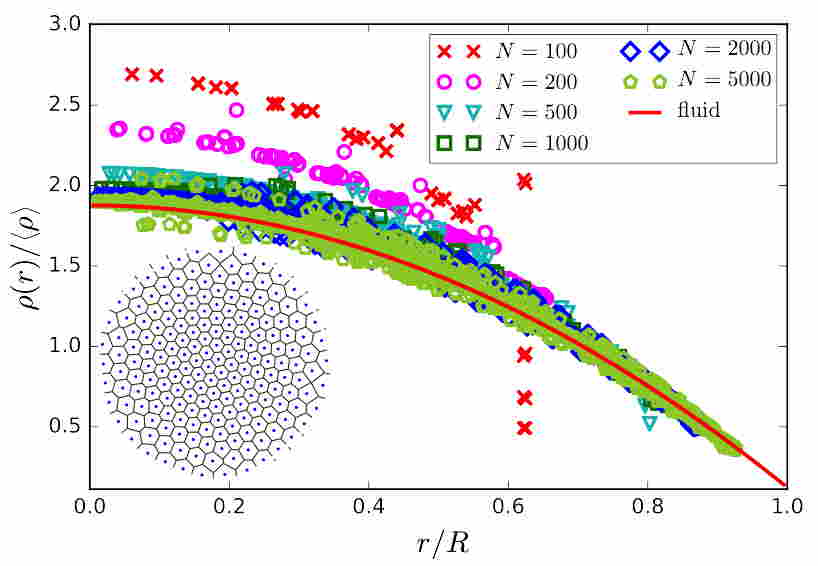}
    \caption{Comparison of local particle density measured from simulated clusters under parabolic confinement with particle number indicated in the legend.  The values correspond to a target reduced size of $\kappa R = 30$, corresponding to the short range interactions.  Here $\langle \rho \rangle$ is the mean density of the cluster. The solid red curve shows the prediction from the fluid density model.  Local density in simulated clusters are derived from the areas per cell of Voronoi tesselation of ground state clusters, as shown for example in the inset for $N=200$.  Note that the areas of the outer most Voronoi cells cannot be defined, and hence points for $r \simeq R$ are omitted from simulated density profiles. }
    \label{fig: parabolicden}
\end{figure} 

This suggests that while fluid density profile may favor very large numbers of positive disclinations for large $\kappa R$, conformal crystals may not be able to achieve such states.  As a consequence, we expect the particle density profiles may depart substantially from the fluid density prediction in the large $\kappa R$ regime.  In Fig. \ref{fig: parabolicden}, we plot the radial profiles local density of simulated clusters (extracted from Voronoi tesselations) for the case of short-range repulsions, $\kappa R = 30$, and compare it the inverted parabolic prediction of $\rho_f(r)$ from  eq. (\ref{eq: rhofn2}).  We note that ultimately for large numbers of particles ($N \gtrsim 500$), density patterns of simulated clusters tend towards fluid density predictions, approaching quantitative agreement only for the largest numbers $N =5000$.  For smaller numbers of particles, densities of the simulated clusters significantly exceed fluid density predictions in the cluster center by large amounts (up to 44\%).

Taken together, this suggests the ability of conformal crystal packings to achieve energetically optimal patterns density (and disclinations) from finite particle numbers is limited, and becomes more limited as $N$ decreases.

\begin{figure*}[th!]
    \centering
    \includegraphics[width=0.98\textwidth]{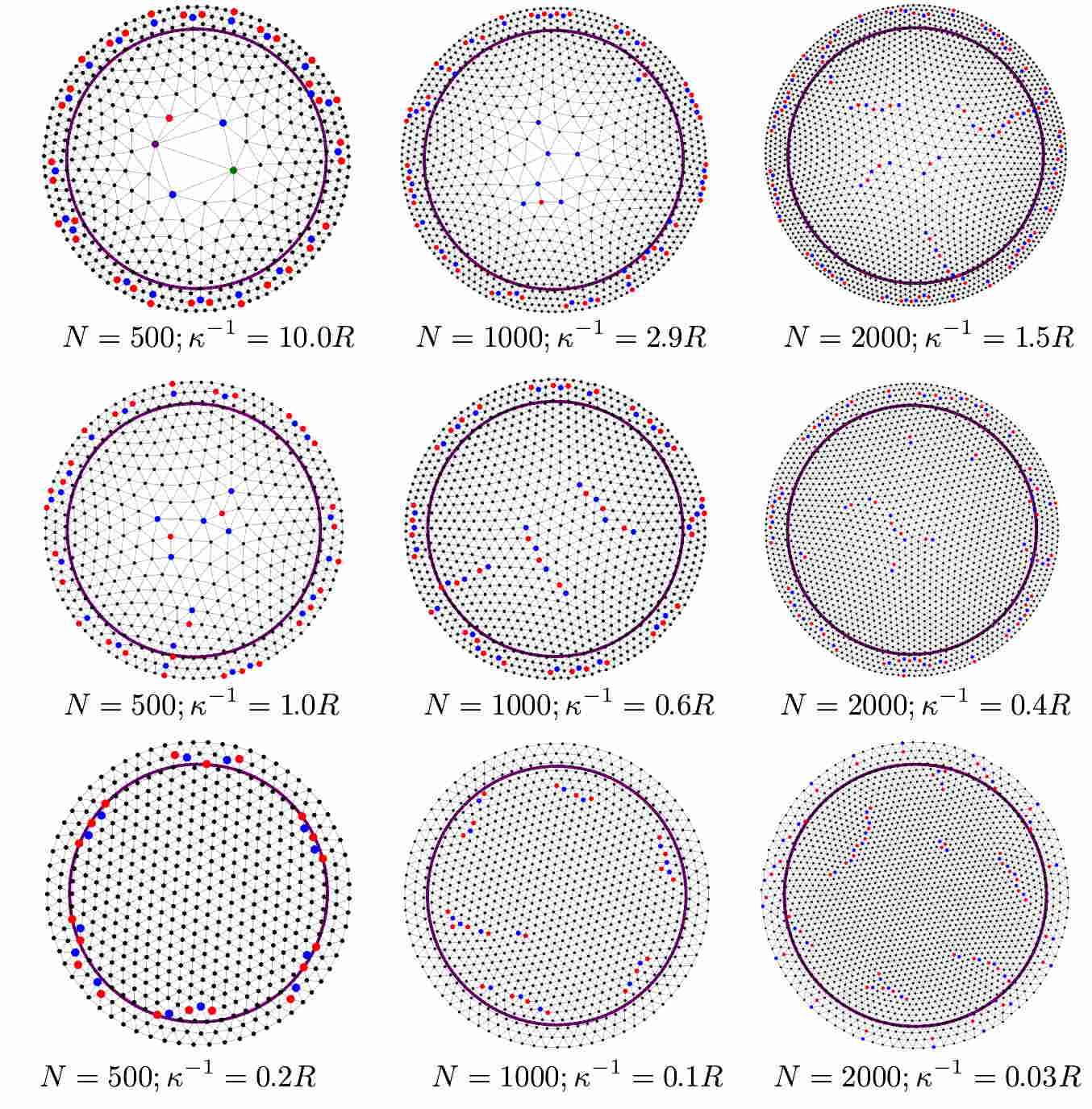}
    \caption{
   Simulated ground state configurations for anharmonically ($n=4$) confined clusters of $N = 500, 1000$ and $2000$ with variable screening lengths.  Delaunay triangulation shows the neighbor network, with site coordination indicated by black (6), red (5), blue (7), green (8) and purple (9).  The purple circle denotes what is used that define the ``interior" region for bulk disclination charge count, $85\%$ of the radial distance to the outer boundary. 
}
    \label{fig:r4snapshot}
\end{figure*}

\subsection{Anharmonic confinement (quartic potential)}

We now turn to the case anharmonic confinement, in the form of a quartic external potential.  For this case, the optimal fluid density profile is described by the fourth order polynomial of radius,
\begin{equation}
\label{eq: rhof4}
\rho_f(r)  = \rho_0\big[ f_4 (\kappa R) + 16 (\kappa r)^2  - (\kappa r)^4 \big]  \ \ \ \ {\rm for \ } n =4 .
\end{equation}
Here, we note that the character of this fluid density profile is strongly dependent on interaction range as shown in Fig. \ref{fig: anharmonicden} below.  For small $\kappa R$ , the profile is dominated by the parabolic increase in density, leading to a {\it convex} pattern with density spread to larger radii.  In the opposite regime $\kappa R \gg 1$, the constant and quartic terms dominate, leading instead to a concave pattern of radially decreasing density.  Like the case of harmonic confinement, here also the fluid density is also predicted to vanish at the cluster edge in the limit of very short-range interactions, $\rho_f(r)  \simeq \rho_0\kappa^4 (R^4 -r^4)$.  The crossover of $\rho_f(r)$  from convex to concave with interaction range, is also reflected in the fluid defect charge and its dependence on interaction range,
\begin{equation}
Q_f = 12 \rho_0 \frac{ (\kappa R)^4 - 8 (\kappa R)^2}{\rho_f(R)}  \ \ \ \ {\rm for \ } n =4 \label{eq:r4QR}.
\end{equation}
For this case, the sign of the fluid defect charge varies with interaction range.  In the limit of long-range interactions, this prediction tends towards a minimal, negative charge, $Q_f( \kappa R \to 0) \to -6$.  It crosses over from net negative to net positive defects at $\kappa R = 2\sqrt{2}$, and in the limit of short range interactions, follows the same positive divergence as the harmonic case, $Q_f (\kappa R \to \infty) \sim \kappa R$.

Fig. \ref{fig:r4snapshot}, shows how simulated ground-state patterns for quartic potentials evolve with interaction range for three values $N$.  Generally speaking, when repulsions are long-ranged, the clusters exhibit low densities at the cores, punctuated with excess negative disclinations (7-, 8- or even 9-fold).  To a large extent, these disclinations are localized towards the center of the cluster. In the opposite, short-ranged repulsive regime, this pattern inverts, with the central region of the cluster showing a (largely uniform) higher density than its periphery.  The excess charge of defects appearing in the interior large $\kappa R$ patterns also flips from negative to positive.

\begin{figure}
    \centering
    \includegraphics[width=0.48\textwidth]{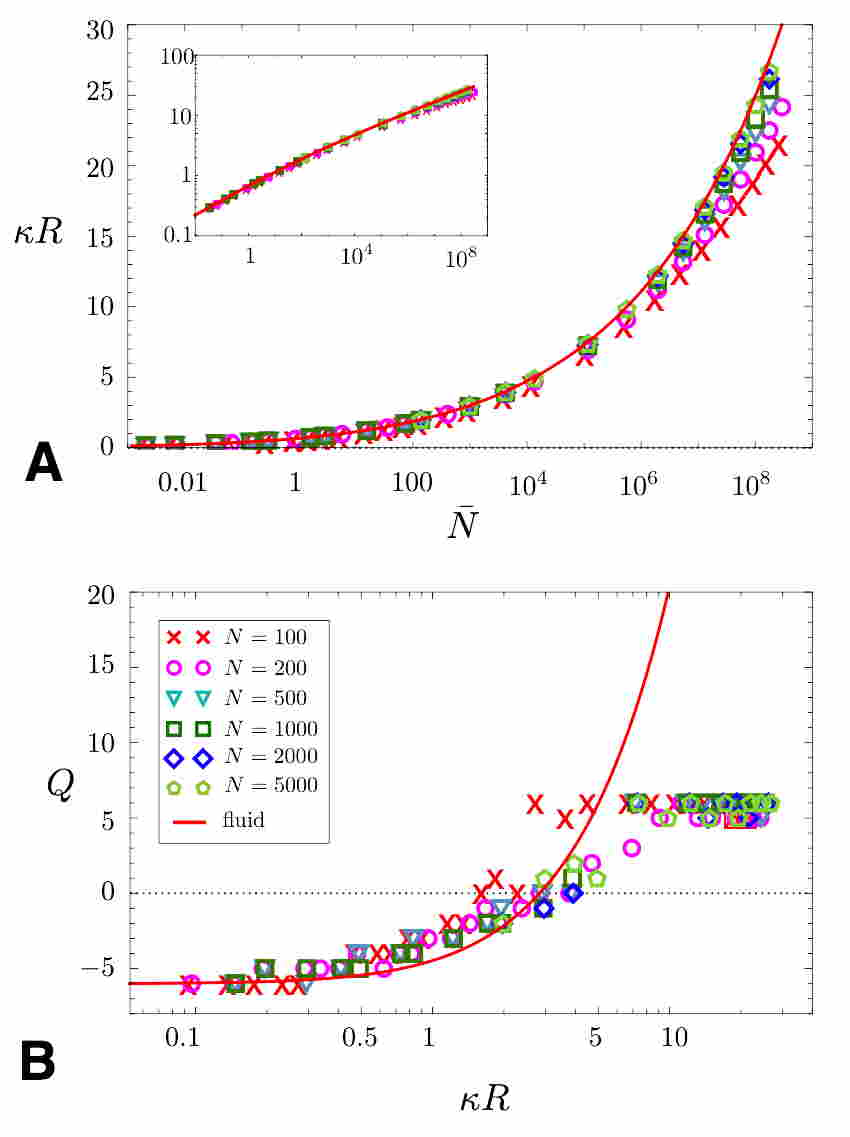}
  \caption{Comparison of size (A) and disclination charge (B) between fluid density model and simulated (discrete) particle ground states for anharmonic confinement ($n=4$).  In (A), comparison of the equation of state in eq. (\ref{eq: kRvsN}) for scaled cluster size vs. scale particle in the fluid density model to the equilibrium cluster size of simulated clusters.  In (B), the predicted disclination charge in the fluid density model compared to interior defect count (within the 85\% of the radial distance from the center), both plotted as functions of the reduced cluster size $\kappa R$.  The numbers of particles in simulate clusters are indicated by the legend in (B).  }
    \label{fig: quarticsim}
\end{figure}

Fig. \ref{fig: quarticsim} compares the cluster size and interior disclination charge of simulated clusters to the fluid density model.  As in the case of harmonic confinement, Fig. \ref{fig: quarticsim}A shows good agreement of the predicted dependence of $\kappa R$ with scaled particle size $\bar{N}$, with the except of modest error for small particle numbers (i.e. $N \lesssim 1000$) in the limit of short range interactions. In Fig. \ref{fig: quarticsim}B, we observe that simulated ground states follow the tendency towards $Q \to -6$ predicted by the fluid density model, and further, the qualitative crossover from net negative to net positive defect states increasing from small to large scaled sizes $\kappa R$.  However, as in the harmonic case, the number of positive defects in the interior of the simulated clusters saturates as a finite value for short range interaction, in disagreement with the unbound growth of $Q_f$ for $\kappa R \gg1$.  

\begin{figure}
    \centering
    \includegraphics[width=0.48\textwidth]{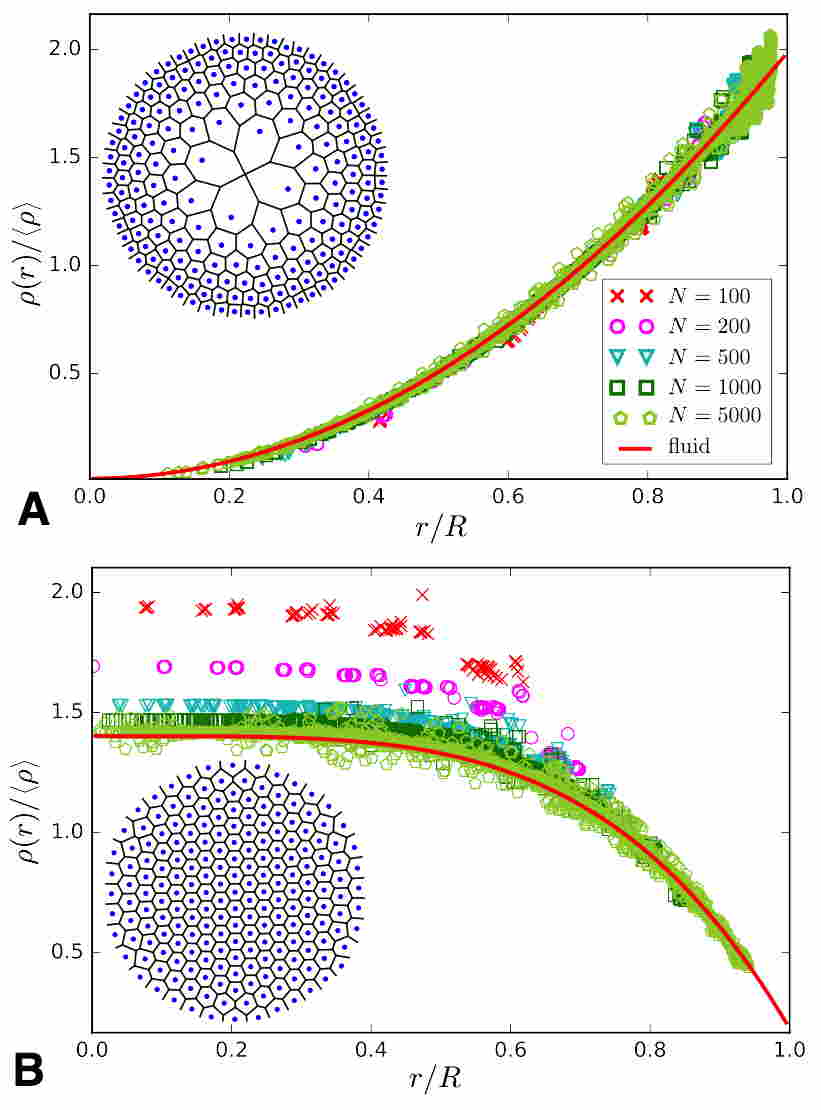}
    \caption{Comparison of local particle density measured from simulated clusters under anharmonic ($n=4$) confinement with particle number indicated in the legend.  The values correspond to a target reduced sizes of $\kappa R = 0.1$ in (A) and  $\kappa R = 30$ (B) corresponding to the long- and short-range interactions, respectively.  The solid red curves shows the predictions from the fluid density model. Here $\langle \rho \rangle$ is the mean density of the cluster.  Insets show example Voronoi tesselations for $N=200$ clusters. }
    \label{fig: anharmonicden}
\end{figure}

The asymmetric agreement between simulations and fluid defect predictions suggests that limitations to approximate optimal fluid density patterns in conformal crystals may be much less restrictive for $Q<0$ patterns for long-range repulsions than for the correspond $Q>0$ patterns in the opposite regime of short-range repulsions.  This interpretation is consistent with the comparison of the density profiles in Fig. \ref{fig: anharmonicden}.  In Fig. \ref{fig: anharmonicden}A, we compare density profiles of the simulated ground states to $\rho_f(r)$ first for long-range interactions, $\kappa^{-1} = 10 R$.  This shows that simulated packings agree well with optimal fluid density patterns, even for small values of $N$.  In contrast, the corresponding comparison in Fig. \ref{fig: anharmonicden}B for short-range repulsion,  $\kappa^{-1}= 0.03 R$, shows (like the harmonic case) that density in simulated packings significantly overshoots the prediction from $\rho_f(r)$ in the center of the clusters for small $N$, and only converges to the predicted pattern for sufficiently large $N$.

In summary, for quartic potential, simulations follow the predictions for the fluid density model to adopt net-negative multi-disclination patterns when repulsions are long ranged, even for fairly modest numbers of particles, while the fluid density model grossly overestimates the total charge of positive disclination ground states, and poorly approximates the density of simulated states for smaller $N$.

\section{\label{sec: discrete}Ground states in the discrete defect model}

The observations from the simulated ground states in the previous section suggest overall that conformal crystals at finite $N$ follow some of the basic trends predicted by the fluid density model, in particular, the evolution from convex to concave density profiles with decreasing interaction range, and the corresponding evolution from net negative to net positive disclination charge.  However, extent of agreement in terms of the total disclination charge and quantitative profiles of particle density vary considerably.  Generally speaking, simulated ground states follow the fluid density model for long-range interactions, where disclinations are absent ($n=2$) or negatively charged ($n=4$), while for short-range interactions the fluid density model grossly overestimates the total charge of positive defects, and poorly models the densities low-$N$ clusters.

\begin{figure*}
    \centering
    \includegraphics[width=0.98\textwidth]{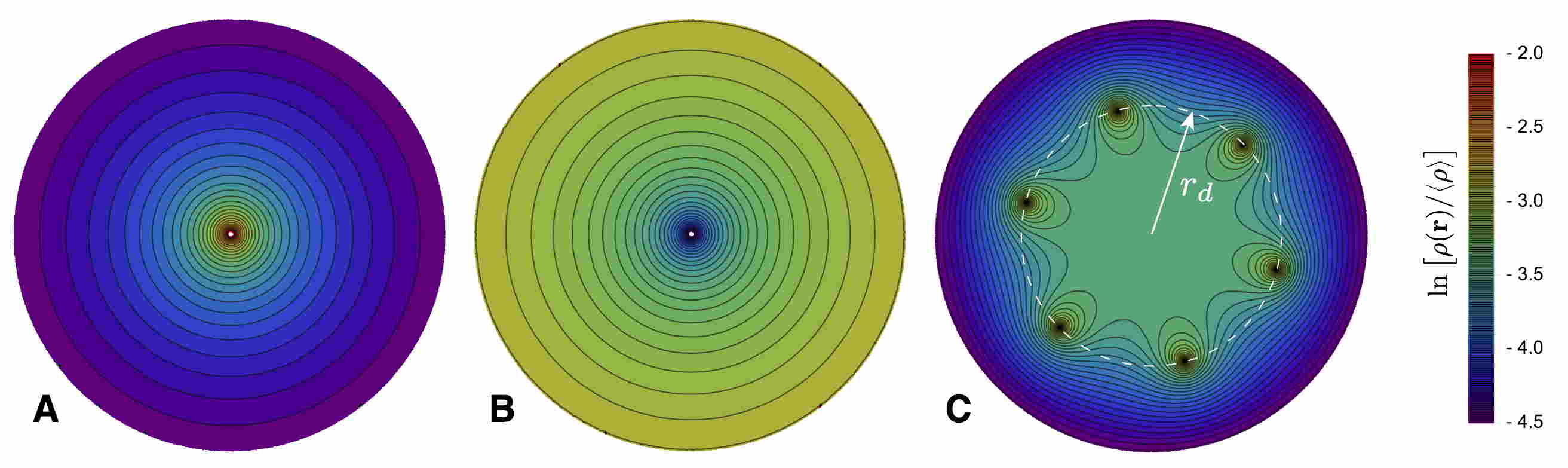}
    \caption{Predicted particle density profiles for conformal crystal with discrete disclinations:  (A) centered $Q=+1$ defect; (B) centered $Q=-1$; and (C) disclination ring of six $Q=+1$ disclinations at radius $r_d$ from the cluster center.  The color profile indicates the local variation of density relative to mean density $\langle \rho \rangle$.}
    \label{fig: defectdensity}
\end{figure*}

To investigate the physical origins of these discrepancies , we turn to the discrete disclination model of conformal crystals introduced in Sec. \ref{sec: discretedefect}.  Specifically, we aim to study the role played by the finite number and discrete nature (i.e. quantization) of disclinations in selecting optimal energy patterns. We first consider the energetics of the axisymmetric patterns, or circular clusters possessing a single disclination (of variable charge) at its center (as in Fig.\ref{fig: defectdensity}A-B) .  We then compare to energetics of distributed multi-disclinations states with density patterns that break the axisymmetry of the confining potential (as in Fig.\ref{fig: defectdensity}C).

\subsection{Centered disclinations}

Here, we consider the energetics monopole disclination patterns within confined 2D screened Coulomb model.  In particular, we consider the axisymmetric density profiles described by eq. (\ref{eq: axidisc}), for a centered disclination of charge $Q$, which is required to be an integer, as opposed to the continuously variable disclination charge implicit to the fluid density model of Sec. \ref{sec: continuum}.  As described in Sec. \ref{sec: discretedefect}, we evaluate the energy for fixed $Q$, $R$, $N$, interaction and confining potential.  We then minimize the results with respect to $R$ to obtain the equilibrium cluster size, and compare the energetics of variable integer disclination charge.

For this case of centered disclinations, the cluster energy can be expressed in a compact form,
\begin{multline}
\frac{E_{cen}}{N(u_0 /\kappa^{n})} = 4 \bar{N} (1-Q/6)^2 \frac{ f_{int}(\kappa R)}{ (\kappa R)^{4(1-Q/6)} }  \\ +2 \frac{ (1-Q/6) }{ n+2 - Q/3} (\kappa R)^n 
\end{multline}
where 
\begin{equation}
f_{int}( t) \equiv \int_0^{t} dy ~K_0(y) y^{1-Q/3}~\int_0^{y} dx ~I_0(x) x^{1-Q/3} ,
\end{equation}
characterizes the pairwise repulsive cost, see eq. (\ref{eq: continuum}), and the scaled particle number is defined as above in eq. (\ref{eq: barN}).  This formulation shows that the  energetics of centered defects is parameterized effectively by three dimensionless quantities, defect charge $Q$, scaled number $\bar{N}$, and ratio of cluster size to screening length $\kappa R$ with the latter quantity being optimized for equilibrium structures.  

\begin{figure}
    \centering
    \includegraphics[width=0.48\textwidth]{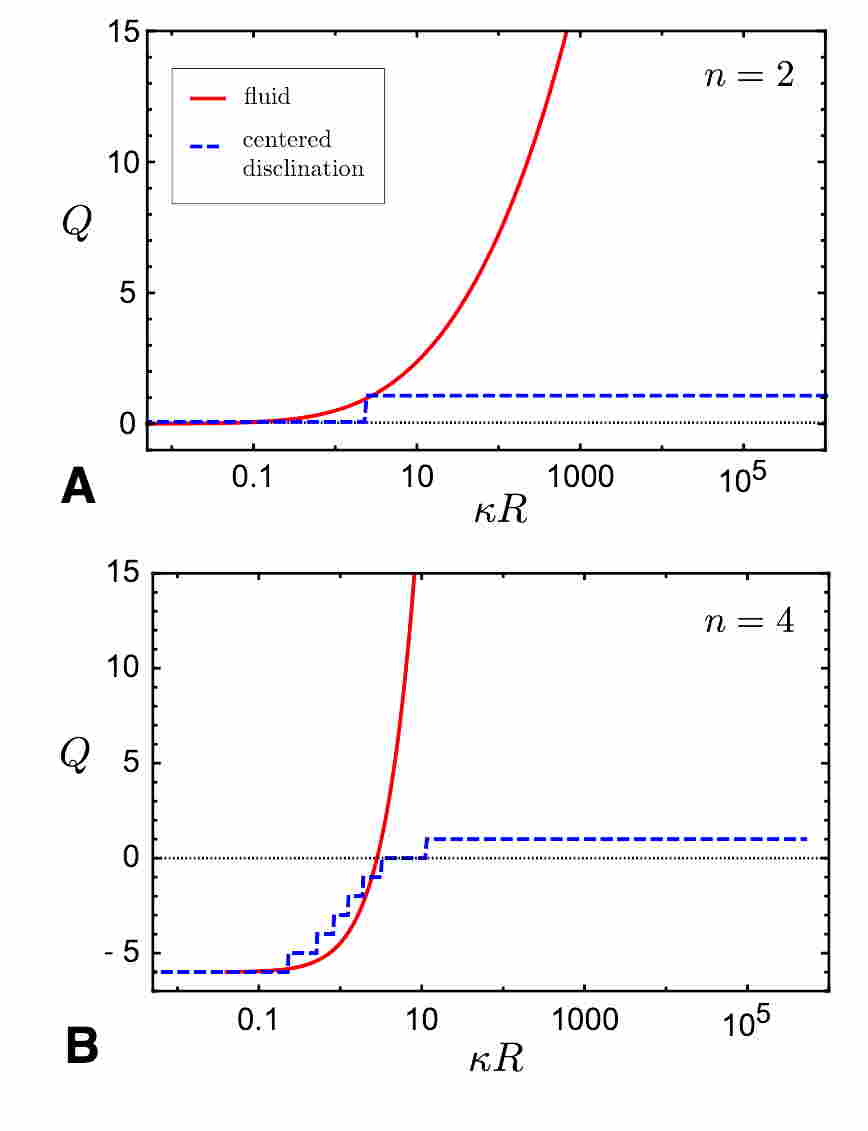}
    \caption{Comparison of energetically favored central disclination conformal crystal (discrete disclination theory) to the predicted defect charge from the fluid density model as function of reduced cluster size $\kappa R$:  (A) harmonic confinement; and (B) anharmonic confinement.}
    \label{fig: centereddisc}
\end{figure}

We consider the energetics of central-disclination configurations with $Q$ ranging from -7 up to +5, for a both confinement potentials, as well as a range of scaled particle numbers from $\bar{N} = 1$ to $10^{32}$ which effectively allows us to span the range of long-range $(\kappa R) \ll1$) to short-ranged $(\kappa R) \gg1$) repulsions.  Fig. \ref{fig: centereddisc}A, shows results for the optimal centered disclination charge for harmonic confinement.  For equilibrium sizes $\kappa R \leq 2.23$, $Q =0$(uniform, defect-free ) states are energetically favored over other axisymmetric patterns.  Above this size, there is a transition to preference for $Q=+1$ among all centered defect states, which persists up to the largest cluster sizes (or shortest interactions ranges) consider $(\kappa R) \gtrsim 10^8$.  

Fig. \ref{fig: centereddisc}B, shows the corresponding prediction for central-disclination patterns for the anharmonic (quartic) confinement.  For this case, the longest interaction ranges (or smallest cluster sizes) adopt an optimal charge of $Q=-6$ that gradually increases in a stepwise fashion to a maximal central charge of $Q =+1$ over the range from $\kappa R =0.23$ to $11.36$.  Hence, both the $n=2$ and $n=4$ cases show a rough agreement with the defect count in simulated ground states {\it and} the predictions of the fluid defect count $Q_f$ in the regime of longer-range interactions, $\kappa R \ll 1$.  However, in the opposite regime $\kappa R \gg 1$, minimal energy central-defect patterns remain capped at $Q=+1$, falling short of the maximal defect charge observed in simulations ($\approx +6$), not to mention the divergent growth as $\kappa R \to \infty$ predicted by the fluid density model.

\subsection{Decentralized disclination rings}

Here we consider influence of discrete disclination arrangements on the ground-state energetics.  Specifically, in contrast to the axisymmetric patterns which assume defect-confinement to the geometric center of the cluster, here we consider defect patterns that are spread into $|Q|$-fold rings sitting at constant radius $0 \leq  r_d \leq R$ from the center, as shown in Fig. \ref{fig: defectdensity} C.  In Appendix \ref{sec:split}, we describe the generalized from of the conformal crystal density patterns and energetics for these $|Q|$-fold rings.  Critically, the ``fission'' of a single, centralized disclination charge allows for a much broader class of density patterns than can be achieved by axisymmetric defects, for examples the $Q = + 6$ state  shown in Fig. \ref{fig: defectdensity} C, which is smooth everywhere but the singular positions of the disclinations.

The energy is computed from eqs. (\ref{eq:EconSplit}) and (\ref{eq:K04Integral}) for a fixed number ( $|Q|$),  charge ($\pm$) of disclinations, scaled particle number $\bar{N}$.  The total energy of each configuration is numerically minimized cluster size $\kappa R$ for a series of variable radii $r_d/R$ for the disclination ring.   

\begin{figure}
    \centering
    \includegraphics[width=0.48\textwidth]{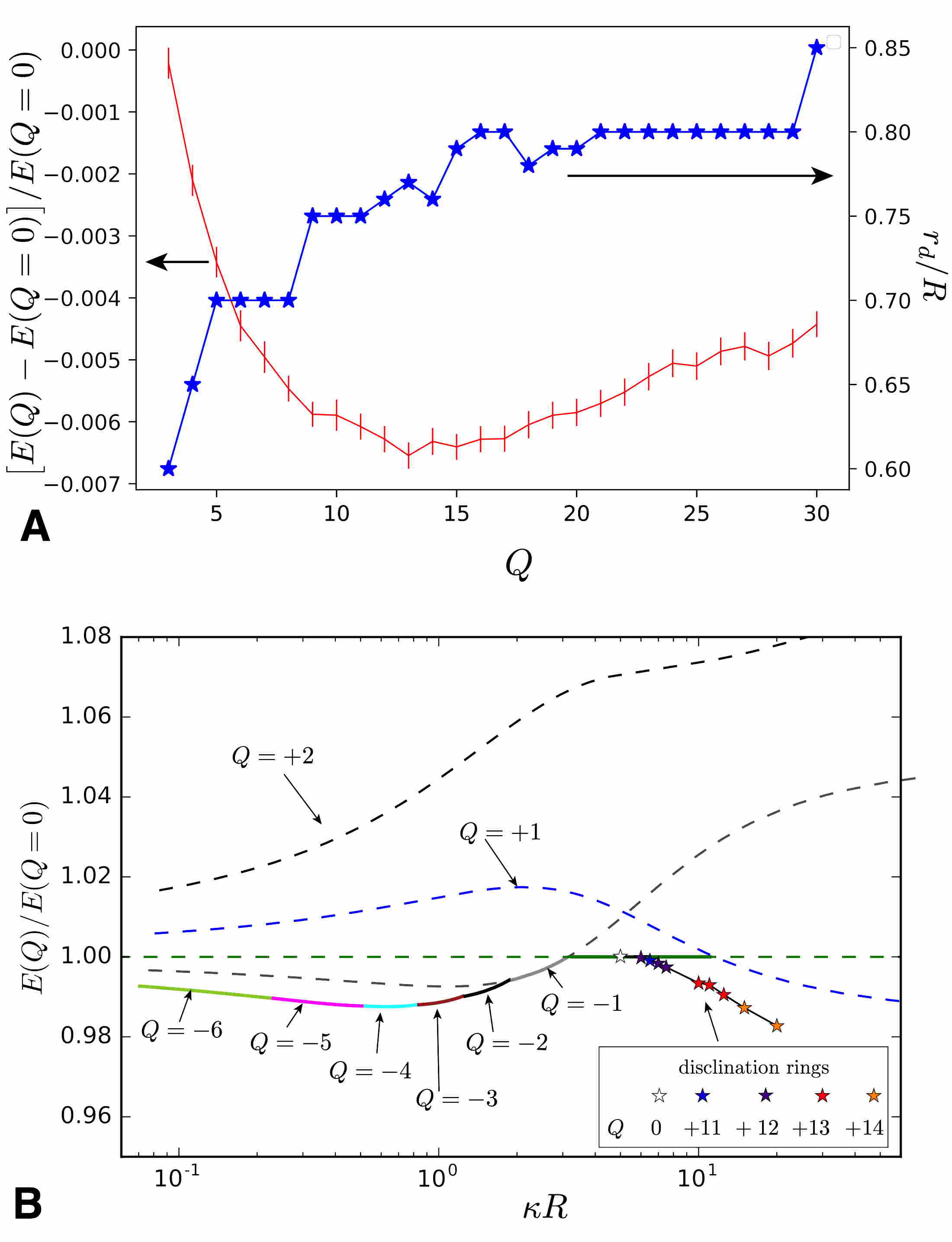}
     \caption{Energetics of multi-disclination ring conformal crystals of anharmonically confined particles.  (A) Energy and optimal radius $r_d$ of as function of variable disclination number $Q$ for with $\bar{N} = 80000$ (or  $\kappa R \simeq 10$).   (B) Lowest energy configuration $E(Q)$ vs $\kappa R$ plot scaled by the uniform energy $E(Q=0)$, with starred points indication energetically optimal disclinations configuration, and the remaining curves denoting centered disclination configurations.
        \label{fig: discring}
    }
\end{figure}

For these nonaxisymmetric and decentralized patterns, we focus only on the case of the quartic potential, as this scenario exhibits a transition from negative charge to positive charge disclinations with increasing $\kappa$.   Here, we consider $|Q|$-fold patterns possessing $2\leq |Q| \leq 40$ symmetrically arrayed defects. Notably, we find that for $Q<0$ states, the optimal pattern also always favors axisymmetric, centered states (as described in the previous section).  It is only for the positive disclination states that decentralized ring patterns achieve lower total energy than central $Q>0$ states.  Fig \ref{fig: discring} A, shows an example of results for $\bar{ N} = 800000$ (corresponding approximately to cluster sizes $\kappa R \approx 10$), plotting the optimal ring radius $r_d$ and energy of the $|Q|$-fold ring as function total $Q$, showing a minimal energy for $Q=+13$ and $r_r = 0.775 R$.

We computed the optimal disclination ring pattern for $\bar{N} = 22400$ to $560000000$, which corresponds to a range of equilibrium cluster sizes $\kappa R \simeq 5.5 - 30$.  The results of the comparison between energetics of centered defect patterns to split-ring patterns in shown in Fig. \ref{fig: discring}B.  Above a threshold value of $\bar{N} \simeq 22400$, we find that split disclination patterns of $Q = +7$ become stable over the axisymmetric patterns, which at this threshold are have lowest energy in defect-free ($Q=0$) configurations.  Hence, the transition from $Q \leq 0$ to positively charged, split-ring configurations preempts the stability of the $Q>1$ {\it centered} disclination patterns, which does not occur until $\bar{N} \simeq 22400$ ($\kappa R \simeq 5.5$).  Going beyond the threshold between neutral and split ring configurations, we find that the optimal charge increases from $Q=+13$ at $\bar{N} \simeq 800000$ ($\kappa R \simeq 10$) and then to $Q=+17$ at $\bar{N} \simeq 176000000$ ($\kappa R \simeq 25$).  This indicates a strong bias for positive disclinations to spread out away from the cluster center and localize much closer to its boundary.  This is consistent with plot of optimal ring radius vs. total disclination charge shown in Fig. \ref{fig: discring}A, which shows that minimal-energy values of $r_d/R$ increase with $Q$.

\begin{figure}
    \centering
    \includegraphics[width=0.48\textwidth]{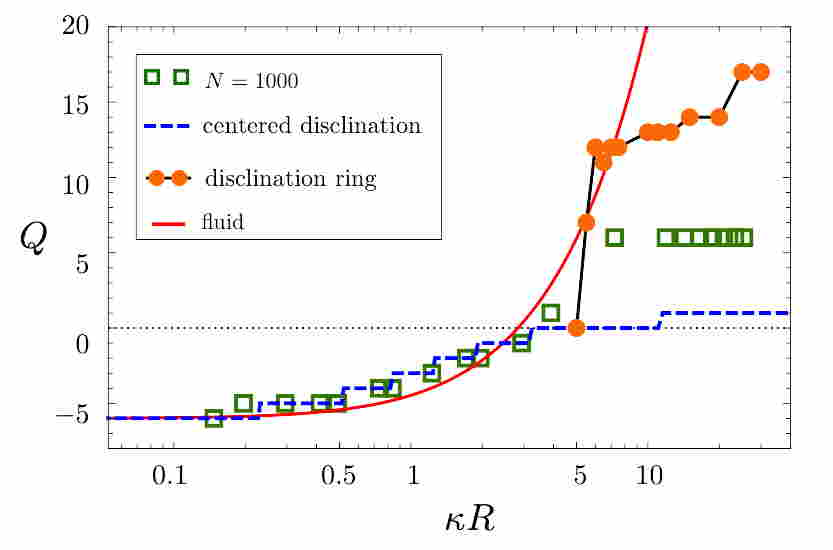}
     \caption{ Comparison disclination charge of the optimal disclination rings to centered disclinations, predicted charge of fluid density model and measured interior defect charge in $N=1000$ simulated clusters for anharmonicially confined ($n=4$) ground states.
     \label{fig: QvskR}
    }
\end{figure}

Finally, we compare the total defect counts of both optimal  split-ring configurations and axisymmetric (centered) configurations of discrete dislcinations to both numerically simulated ground states and the fluid-density model predictions for quartic confinement in Fig. \ref{fig: QvskR}.  We note that the split defect configurations share the feature with the simulated ground states in the short-interaction range that the total disclination charge becomes positive, but remains relatively constant in comparison to the divergent prediction of the fluid density model.  For example, for largest cluster size studied $\kappa R \simeq 30$ we find an optimal charge of +17 split disclinations, which while larger than the $+6$ charge measured from the simulated clusters falls more than an order of magnitude below the fluid density prediction of $Q_f(\kappa R = 30) \simeq 80$.   In the following section we discuss the comparison between these three different predictions for optimal defect patterns.



\section{Discussion}
\label{sec: discussion}

We have compared two distinct theoretical models for optimal disclinations patterns in conformal crystal ground states of confined 2D clusters of particles with screened Coulomb repulsions.  The fluid-density model provides analytic predictions for energetically favored pattern of particle density in the cluster.  Applying the conditions of geometric compatibility between bond order and disclinations in conformal crystal, eq. (\ref{eq: conformal}), but neglecting constraints of localization and quantization of defect charge, yields a prediction, eq. (\ref{eq: fluiddefect}) for the {\it continuous} fluid defect charge density $s_f(r)$ corresponding to the energetically optimal particle density, which is analogous to the Nye picture of continuous defects in bent crystals.  Consistent with previous studies that adopt this approach~\cite{mughal_topological_2007,yao_topological_2013,Soni2018}, we find that this fluid-defect model works well to captures the basic trend increasing of net disclination charge, and the corresponding crossover from convex to concave radial profiles of density in the cluster, with the ratio of equilibrium cluster size to screening length, $\kappa R$.  While this model captures fairly well the quantitative features of optimal clusters in the limit of long-range interactions $\kappa R \ll 1$ where net disclination charge is non-positive, we find that it dramatically overpredicts the total charge of positive disclination ground states in the opposite limit of short range interactions, $\kappa R \gg 1$.  While $Q_f( \kappa R \to \infty) \to \infty$, simulated ground state clusters exhibit a {\it bounded increase} in the total disclination charge that saturates for $\kappa R \gtrsim 5$.  Notably, we observe also a strongly $N$-dependent mismatch between the predicted fluid density pattern and simulated clusters in this short-range interaction regime.  For $N \lesssim 10^3$, simulated clusters overshoot the predictions of particle density from the fluid model in the core of the clusters, while no significant discrepancy is observed in the long-range interaction regime, where defect patterns are non-positive in charge.

A second theoretical approach, akin to the Taylor theory of solid dislocations, treats defects as quantized and monopole sources of conformal distortion, according to the compatibility relation in eq. (\ref{eq: conformal}).  Consideration of the energetics of the resulting density patterns that can be achieved for various arrangements of elementary $Q=\pm1$ disclinations, then provides a distinct set of predictions for the dependence of defect pattern on interaction range.  Largely speaking, the discrete defect theories tend to agree well with both the simulations and fluid density predictions in the long-range interaction regime, where disclinations patterns are non-positive in total charge.  As described in Sec. \ref{sec: continuum}, the fluid density model predicts that optimal defect patterns tend towards a central delta-function in the $\kappa R \to 0$ limit.  Correspondingly, we find that central patterns of discrete negative disclinations do well to capture the observed increase in $Q$ with cluster size in the $\kappa R \ll 1$ regime.  Central patterns of discrete disclinations, however, fail to capture the observed growth in $Q$ into the short-range interaction regime, $\kappa R \gg 1$.  Notably, the optimal charge of the axisymmetric disclination pattern never exceeds $Q = +1$ even in the limit of $\kappa R \to \infty$.  Instead, we find that optimal patterns of discrete positive disclinations break axisymmetry, spreading into rings that are situated away from the cluster center and tend towards the its boundary with increased total charge.  Upon taking into account this sensitivity of $Q>0$ ground states to discrete defect position, we predict a far more modest increase of total defect charge for $\kappa R  \gg 1$ than the fluid density model, in better qualitative agreement with numerical simulation results.

Taken together, these results suggest a marked asymmetry between the function of positive and negative disclinations in ground states of this class of models.  This is unlike the more familiar picture of disclinations canonical crystals, i.e. formed by short-range cohesive bonding, which is based on linearized continuum elasticity theory~\cite{nelson_defects_2002}.  In that standard description, the induced stress, elastic interactions and coupling to curvature simply flip sign as $Q \to -Q$~\cite{seung_defects_1988}.  What then accounts for the asymmetric behavior we find here for conformal crystal groundstates with respect to disclination charge?

\begin{figure}[t]
    \centering
    \includegraphics[width=0.48\textwidth]{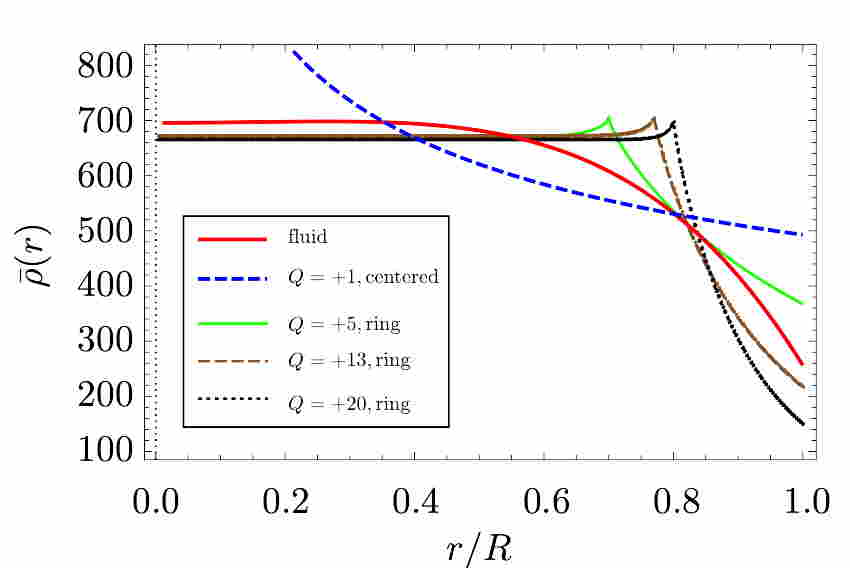}
    \caption{Comparison of the optimal fluid density profile to the radially-averaged density $\bar{\rho}(r)$ profiles predicted for different discrete defect patterns in anharmonically confined clusters with the same parameters as shown in Fig. \ref{fig: discring}.  For disclination ring patterns, the state with the energetically optimal $r_d$ is shown.
        \label{fig:rhoSplitCompare}
    }
\end{figure}

In part, the distinction derives from energetic preferences for defects locations and its dependence on the range of interaction.  As shown in Sec. \ref{sec: continuum}, the optimal patterns fluid density in the long-range interaction (purely Coulomb) limit corresponds to central, delta-function distributions of disclination charge.  That is, the optimal patterns of particle density correspond directly to localized point sources of disclination charge.  In contrast, in the opposite limit of short-range repulsion, the optimal patterns of the fluid density model predicts continuous disclination diverge that diverge cluster edges.  Hence, as interactions become more screened, optimal density patterns favor delocalized defect distributions that are not compatible with their discretized, point-like nature of disclinations in the bond order field, and multi-defect patterns lead to suboptimal density patterns. 

Spreading of multiple positive disclinations into rings considerably improves their ability to approximate the optimal fluid density patterns.  For example, see the radially averaged density profiles for anharmonically-confined, short-range repulsive particles ($\bar{N} = 800000$, corresponding to $\kappa R \approx 10$) shown in Fig. \ref{fig:rhoSplitCompare}.  The optimal fluid density pattern (shown as red) is concave and according to the integrated fluid defect charge would correspond to $Q_f \simeq 20$.  The dashed blue line shows the density profile of single centered $Q=+1$ defect, which has the same sign of topological charge, but its singular, convex profile poorly approximates the energetically preferred one.  In comparison, we also show radial averages of density in split disclination rings, which evidently better approximates the optimal fluid density, with a roughly constant density core region that tapers between the defect ring and diminished density at the outer edge.

\begin{figure}[t]
    \centering
    \includegraphics[width=0.48\textwidth]{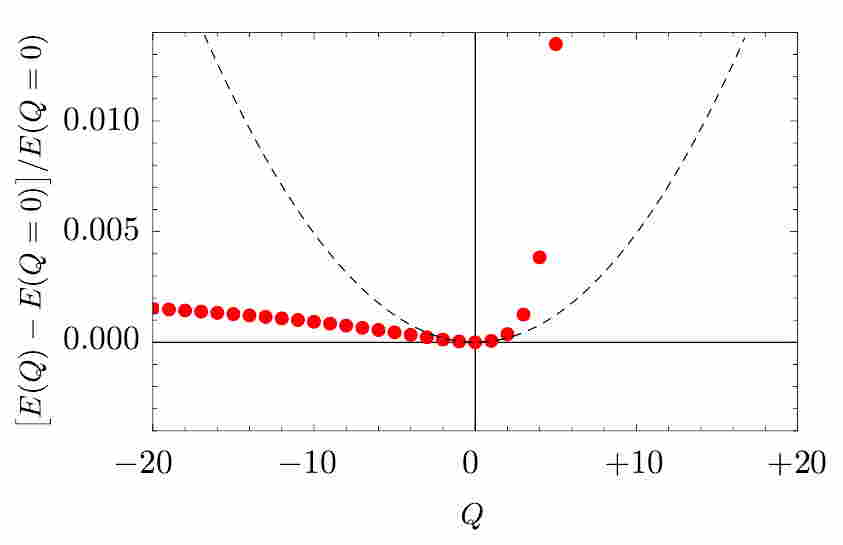}
    \caption{Dependence of ``self-energy'' of single disclinations on defect charge $Q$ for center defects harmonically confined ($n=2$) clusters with long-range interactions ($\kappa \to \infty$).  Red dots show predictions from the conformal crystal (i.e. discrete defect) theory, and the dashed line shows the parabolic fit to the small $Q$ limit, highlighting a large asymmetry for $Q \to -Q$.  Note from eq. (\ref{eq: rhofn2}) that the optimal fluid density for $\kappa \to 0$ is uniform, and hence, all $Q\neq 0$ states represent excitations of the ground state.     \label{fig:r2lnrNPlog}
    }
\end{figure}

Notably the fact that optimal location of discrete dislocations gets pushed to the outer edge for $\kappa R \gg 1$ likely also accounts for some of the saturation effects observed in simulated clusters.  That is, precisely where the preferred total defect count is growing large, the optimal position approaches arbitrarily close the free boundary of the cluster.  This, in combination with the presences of topologically-necessary disclinations on the boundary, compounds the difficulty to count such defects as ``interior" to the packing (see Appendix \ref{app:countQ}).

A second ingredient to the distinct behavior of positive and negative disclinations is the asymmetric dependence of conformal crystal structure on the {\it sign} of topological charge.  As described in Sec. \ref{sec: discretedefect}, disclinations act monopole sources for conformal distortions of $\ln \rho(\rv)$, and as such lead to near-field singularity at the location of disclinations, $\rho(\rv) \sim  |\rv -\rv_d|^{-Q/3}$, where $\rv_d$ is the defect position, as in eq. (\ref{eq: axidisc}).  While this density pattern is singular for both signs of $Q$, its value is bounded for negative charge but {\it divergent} for positive charges.  This leads to a strong effect on the asymmetry of the ``self-energy'' associated with near-field repulsive cost of locally under- or over-packing the region around the disclination core.  For example, in Fig. \ref{fig:r2lnrNPlog}, see the energy to form a single, centered disclination (relative to the defect-free cluster) vs. topological charge $Q$ in a harmonically-confined array of particles with 2D Coulomb repulsion (i.e. $\kappa \to 0$). Due to the asymmetry between local density, the self-energy deviates considerably from the symmetric cost $\propto Q^2$ expected from continuum theory of cohesive crystals~\cite{nelson_defects_2002}.  Note that the energy to form a $Q \to -\infty$ disclination remains finite, while the corresponding energy to form positive disclination diverges at a finite value, $Q \to + 6$.  This result implies that energy cost of concentrate multiple $Q<0$ disclinations is modest, while there is strong incentive to split and separate $Q>0$ disclinations into distant regions of a conformal crystal.  This is consistent with the observation that positive disclination states are only stable in split-ring configurations in our discrete defect calculations, while axisymmetric patterns with $Q<-1$ are generically stable.

The asymmetric sensitivity of ground state energetics to {\it sign} of disclination charge raises another question about the detailed structure of disclinations.  Defect ground states of frustrated cohesive crystals are known to exhibit complex extend chains of alternating sign disclinations, known as scars~\cite{Bowick2000, Irvine2010,azadi_emergent_2014, Azadi2016}.  This is considered a mechanism to spread out the stress of point-like disclinations over larger area, leading to lower elastic energy as the ratio of the macroscopic crystal size grows large compared to the lattice spacing.  Similarly, we observed scar-like chains of $Q= \pm 1$ defects in many of our simulated ground states, particularly as $N$ grows large.  Understanding how the such configurations spread the singular density gradients in conformal crystals and how scarred morphologies may or may not better approximate the energetically optimal fluid density patterns remains an open challenge.

\section{Conclusions}

In summary, despite the strong geometric analogy between curvature frustration in canonical (cohesive) crystals and conformal crystal ground states of externally confined patterns, the heuristic picture of ``curvature screening'' by disclinations in the former class of systems does not simply extend by analogy to physical predictions of ground state structure for the latter.  An open challenge remains how to bridge these two classes of geometrically frustrated systems.  One approach may be to study a generalized continuum elastic description that incorporates the physical ingredients needed to generate conformal crystal ground states (i.e. body forces from the external potential and long-range intra-cluster interactions), while at the same time parameterizing the preferences for isotropic, locally crystalline correlations (i.e. local shear and bulk elasticity).  The compatibility conditions of such a description may provide a more direct and intuitive means to understand the emergent asymmetry in the sensitivity of ground state structure to disclination charge.

\section{Acknowledgments}
The authors are grateful to acknowledge L. Cajamarca, C. Duque and B. Chen for useful discussions.  This work was supported by the National Science Foundation under grant numbers DMR 1608862.  Numerical simulations were performed on the UMass Shared Cluster at the Massachusetts Green High Performance Computing Center. 


\appendix

\section{\label{app:proof}Compatibility of density and disclinations in conformal crystals}

Here we derive the compatibility between disclinations in the bond order field and gradients of the density in a conformal crystals. Following, the original description by Rothen, Pieranski, Rivier and Joyet~\cite{Rothen_1993}, it is particularly convenient to adopt the framework of complex analysis (e.g. \cite{Needham2002}).  In particular, we construct conformal crystals as a mapping from a planar, equitriangular lattice in the complex plane $z_0 = x_0 + i y_0$ to a conformally distorted version of the lattice with coordinates $z=x+iy$.  This map is described by the complex function $z(z_0)$, where $z_0$ is the coordinates in the undistorted reference lattice.   Assuming this map to be analytic everywhere by singular points (i.e. defects), a differential element of the reference states $dz_0 = dx_0 + i dy_0$ (e.g. a neighbor bond in the lattice) is mapped to the element
\begin{multline}
dz = dx + i dy  =\Big(\frac{d z} {dz_0}\Big) dz_0 \\ = ( \xi ~dx_0 - \eta ~dy_0) + i ( \eta~ dx_0+ \xi ~dy_0) 
\label{eq: zmap}
\end{multline}
where $\xi \equiv \partial x/ \partial x_0 = \partial y/ \partial y_0$ and $\eta \equiv -\partial x/ \partial y_0 = \partial y/ \partial x_0$, according the Cauchy-Riemann conditions.  Hence, the function describing the conformal distortion can be written in the form $d z/dz_0=\Omega(z) e^{i \theta (z)}$, i.e. $\xi = \Omega(z)  \cos \theta(z)$ and $\eta = \Omega(z) \sin \theta(z)$, and eq. (\ref{eq: zmap}) can be written as a matrix transform
\begin{equation}
 \left( \begin{array}{c} dx \\ dy  \end{array} \right)= \Omega  \left( \begin{array}{c c} \cos \theta & -\sin \theta \\ \sin\theta & \cos \theta \end{array} \right) \left( \begin{array}{c} dx_0 \\ dy_0  \end{array} \right) ,
\end{equation}
which clarifies that an analytic map corresponds to rigid rotation of $dz_0$ by $\theta(z)$ and isotropic scaling of length by the {\it conformal factor} $\Omega(z)$.  Because $z(z_0)$ is analytic, the function $(d z/dz_0)= \xi (z) + i \eta (z)$ is also an analytic function of $z$.  Applying the Cauchy-Riemann conditions to this function (i.e. $\partial \xi/ \partial x = \partial \eta/ \partial y$ and $ \partial \xi/ \partial y = -\partial \eta/ \partial x$)
we have,
\begin{eqnarray}
\nonumber
\partial_x \big( \Omega \cos \theta \big) &=& \partial_y \big( \Omega \sin \theta \big) \\
\partial_y \big( \Omega \cos \theta \big) &=& -\partial_x \big( \Omega \sin \theta \big)
\end{eqnarray}
or
\begin{eqnarray}
\nonumber
\partial_x \theta  &=&\Omega^{-1} \partial_y\Omega  \\
\partial_y \theta  &=&\Omega^{-1} \partial_x\Omega .
\end{eqnarray}
Using this result and the fact that density of the conformal crystal $\rho (\rv) = \Omega^{-2}  \rho_0$, where $\rho_0$ is the uniform density of the equitriangular reference state, we have the relation between row curvature and density gradients in eq. (\ref{eq: benddensity}).

\section{\label{app:K0Cal} Fluid density of confined clusters with 2D screened Coulomb repulsions}

To solve mean-field equations, eqs. (\ref{eq: number}) and (\ref{eq: mu}), for the optimal fluid density pattern we introduce the potential function $\phi(r)$ as
\begin{equation}
\phi(\mathbf{r}) \equiv \int \rho_f(\mathbf{r'}) v_{int}(|\mathbf{r}-\mathbf{r'}|) d \mathbf{r'} \label{eq:PhiDefinition} ,
\end{equation}
so that stationary eq.(\ref{eq: mu}) becomes
\begin{eqnarray}
\phi(\mathbf{r})  =  \mu - U(\mathbf{r}) \label{eq:phiEQ}
\end{eqnarray}
Using the pair wise interactions, form eq. (\ref{eq: Vint}) we have
\begin{eqnarray}
\phi(r) = v_{0} \int_{0}^{R}  dr \rho_f(r') dr' v (r,r')
\label{eq:phiK0} 
\end{eqnarray}
where
\begin{eqnarray}
v(r,r') =  r '\int_{0}^{2 \pi} d \theta K_{0}(\kappa \sqrt{r^2+r'^2-2r r' \cos \theta}) \nonumber \\
\end{eqnarray}
Using the  Bessel function indentify \cite{arfken_mathematical_1985}:
\begin{eqnarray}
\int\limits_{0}^{\pi}  K_{0}(\sqrt{a^2+b^2-2ab \cos x })  d x = \pi \times \begin{cases}
 I_{0}(a)K_{0}(b), & a < b \\ 
 I_{0}(b)K_{0}(a), & a > b \\
\end{cases} \label{eq:K0} \nonumber\\
\end{eqnarray}
yields 
\begin{eqnarray}
v(r,r') = 2 \pi \times 
\begin{cases}
 I_0(\kappa r') K_0(\kappa r), & r' < r \\ 
 K_0(\kappa r') I_0(\kappa r), & r' \geq r 
 
\end{cases} \label{eq:K0solution}
\end{eqnarray}
from which we get the $\phi(r)$ is
\begin{eqnarray}
\phi(r) &=&v_0 K_0(\kappa r) \int_{0}^{r} d r'  \rho_f(r') {\cal I}(r') \nonumber \\
&+& v_0 I_0(\kappa r) \int_{r}^{R} d r' \rho_f(r') {\cal K}(r') \label{eq:phiSplit}
\end{eqnarray}
where ${\cal I}$ and ${\cal K}$ are
\begin{eqnarray}
{\cal I}(r) &\equiv & 2 \pi I_0(\kappa r) r \\
{\cal K}(r) &\equiv & 2 \pi K_0(\kappa r) r
\end{eqnarray}

To solve Eq. (\ref{eq:phiSplit}), we can introduce a new function $\psi(r)$, 
\begin{eqnarray}
\frac{1}{2} \phi(r) + \psi(r) &=&v_{0} K_0(\kappa r) \int_{0}^{r} d r' \rho_f(r') {\cal I}(r') \label{eq:introPsi1} \\
\frac{1}{2} \phi(r) - \psi(r) &=&v_{0} I_0(\kappa r) \int_{r}^{R} d r' \rho_f(r')  {\cal K}(r') \label{eq:introPsi2}
\end{eqnarray}
Differentiating both sides of eq. (\ref{eq:introPsi1}) and (\ref{eq:introPsi2}) with respect to $r$ yeilds
\begin{eqnarray}
\rho_f(r){\cal I}(r) &=& \frac{d}{d r}[\frac{\frac{1}{2} \phi(r) + \psi(r)}{v_{0} K_0(\kappa r)}] \label{eq:rhoPsi1}\\
-\rho_f(r){\cal K}(r)&=& \frac{d}{d r}[\frac{\frac{1}{2} \phi(r) - \psi(r)}{v_{0} I_0(\kappa r)}] \label{eq:rhoPsi2}
\end{eqnarray}
Combining these to eliminate $\rho(r)$ yields, and inhomogeneous, first0order equation for $\psi(r)$,
\begin{eqnarray}
 {\cal I}(r) \frac{d}{d r}\Big[\frac{\frac{1}{2} \phi(r) - \psi(r)}{v_{0} I_0(\kappa r)}\Big] +  {\cal K}(r) \frac{d}{d r}\Big[\frac{\frac{1}{2} \phi(r) + \psi(r)}{v_{0} K_0(\kappa r)}\Big] = 0 \nonumber\\ \label{eq:noRho}
\end{eqnarray}
which takes the follow from
\begin{eqnarray}
p_1(r) \psi^{'}(r) + p_2(r) \psi(r) + p_3(r) = 0 \label{eq:PsiMain} ;
\end{eqnarray}
where
\begin{eqnarray}
p_1(r) &=& 0\\
p_2(r) &=& \frac{2 \pi}{I_0(\kappa r)K_0(\kappa r)}\\
p_3(r) &=& {\cal I}(r) \Big[\frac{\kappa}{2} \frac{I_0(\kappa r)K_1(\kappa r)-I_1(\kappa r)K_0(\kappa r)} {I_0(\kappa r)^2 K_0(\kappa r)} \phi(r)  \nonumber \\
&+& \frac{\phi'(r)}{I_0(\kappa r)} \Big] ,
\end{eqnarray}
Note that $p_2(r)\geq 0$.

Solving for $\psi(r)$ requires a boundary condition, which is obtained from eq. (\ref{eq:introPsi2}) in the limit of $r \to R$, since the integrand on the right-hand side vanishes,
\begin{equation}
\frac{1}{2} \phi(R) = \psi(R) \label{eq:bdy}
\end{equation}
From this and eq. (\ref{eq:PsiMain}), we have
\begin{eqnarray}
\psi(r) &=& \frac{r}{2} \{ \kappa[I_1(\kappa r)K_0(\kappa r)-I_0(\kappa r)K_1(\kappa r)]\phi(r) \nonumber\\
&-& 2I_0(\kappa r)K_0(\kappa r)\phi'(r) \} \label{eq:PsiSolution}
\end{eqnarray}
Inserting eq. (\ref{eq:PsiSolution}) into eq. (\ref{eq:rhoPsi1}) or (\ref{eq:rhoPsi2}), yields fluid the general form of the density distribution in terms of the unknown potential field
\begin{eqnarray}
\rho_f(r) &=& \frac{1}{2 \pi v_{0}} [\kappa^{2} \phi(r) - \frac{\phi'(r)}{r} - \phi''(r)] \label{eq:rhoPhi} ,
\end{eqnarray}
which is essentially the linearized 2D Poisson-Boltzmann equation for axisymmetric charge density.

Inserting eq. (\ref{eq:phiEQ}) into Eq. (\ref{eq:rhoPhi}), we can get the fluid density distribution in terms of a general external potential, give be eq.  (\ref{eq:rhoGeneralFinal}). Then from the constraint on  particle number eq. (\ref{eq: number}), we have the chemical potential $\mu$ in Eq. (\ref{eq: muR}).

\section{\label{app:countQ} Interior disclination count in simulated ground states}

Here we describe our method to count the topological defects in simulated 2D ground-state clusters.  As discussed in Sec. \ref{sec: simulation}, the total topological defects (interior + boundary) are always 6 due to Euler's theorem (see ~\cite{Bowick2009}). We separate topological defects as two types: inner topological defects $Q_{inner}$ and boundary defects $Q_{bdy}$. 
$Q_{inner}$ represents the total topological charge of the system. $Q_{bdy}$ was used to satisfy the Euler's theorem and resolve the conflict between the locally triangular arrangement of particles and the circular symmetry of the confinement. 

Our goal is to determine a method to count the total interior defect charge, as it is only interior disclinations that give rise to gradients in local areal density in our discrete theory (analogous to the free boundary screening of topological defect in 2D elastic crystals~\cite{Grason2012}).  As a practical matter it is not clear if this boundary layer should be 1, 2 or more particular layers thick, as the circularity of the cluster boundary tend to generically push whatever boundary disclinations are needed at least one layer into the cluster.  Hence, a rigorously motivated and generically-application definition is likely beyond reach.  Instead, we opt for a simple and practical definition, that $Q_{inner}$ simply counts all disclinations within a {\it fixed fraction}, 85\%, of radial distance from the potential minimum (see purple circles in Figs. \ref{fig:r2snapshot} and  \ref{fig:r4snapshot}), and the defects in the remaining outer annulus are classified as ``boundary'' type.  

\begin{figure*}[th!]
    \includegraphics[width=0.9\textwidth]{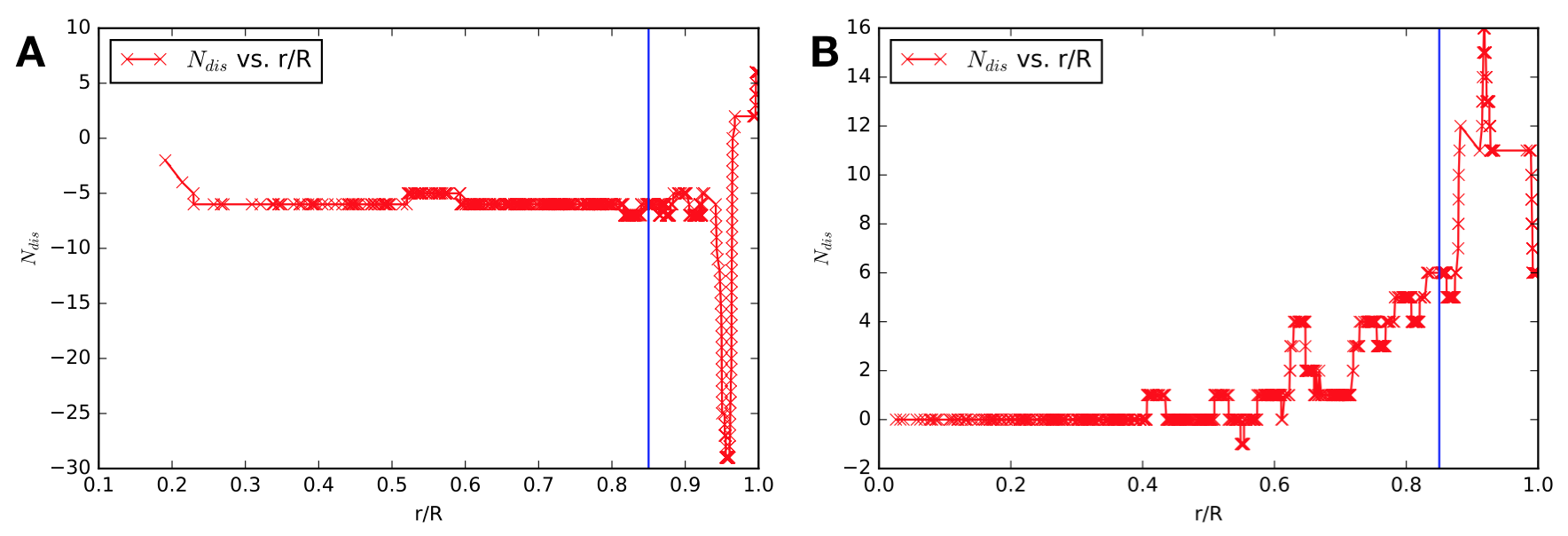}
    \caption{Under quartic confinement, the total disclination charges $N_{dis}(r)$ within a radial fraction $\frac{r}{R}$ for $N = 1000$ at $\kappa^{-1} = 10.0R, 0.03R$. The blue line shows the $r/R = 85\%$ definition .
    }
    \label{fig:QrdivR}
\end{figure*} 

See for example Fig. \ref{fig:QrdivR} which shows the accumulation disclination charges $N_{dis}(r)$ within radius fraction $\frac{r}{R}$ for $N = 1000$ at $\kappa^{-1} = 10.0R, 0.03R$ (their snapshots are in Fig. \ref{fig:r4snapshot}).  As described in the main text, results in for interior charge count are based on a radial threshold of $r/R = 0.85$ (shown as the blue line in Fig. \ref{fig:QrdivR}).

\section{\label{sec:split}Multi-disclination rings}

Here we give the details of the calculation of multi-disclination configurations from the discrete defect model of conformal crystals introducted in Sec. \ref{sec: discretedefect}.  For $M$ topological defects in the cluster, the disclination density has the general form, 
\begin{eqnarray}
s(r) =  q \sum\limits_{\alpha =1}^{M } \delta(\mathbf{r}-\mathbf{r}_\alpha) \label{eq:sigmaM}
\end{eqnarray}
where $q = \pm \frac{\pi}{3}$ is the topological charge per disclination.  From this and eqs. (\ref{eq: psi}) - (\ref{eq: deltapsi}) for a $|Q|$-fold ring of $\pm 1$ charge disclinations at radius $r_d$, we have 
\begin{multline}
\delta \psi(r, \theta) = \pm \frac{1}{12} \sum\limits_{n=1}^{|Q|} \ln\bigg[\frac{R^2+r^2 r_d^2/R^2-2r r_d \cos( \theta - \frac{2 \pi n }{|Q|})}{r^2+r_d^2-2r r_d \cos( \theta - \frac{2 \pi n }{|Q|})} \bigg] .
\end{multline}
From this the density follow
\begin{eqnarray}
\rho(r, \theta) &=&\rho_R H(r/R, r_d/R,  \theta) \\
H(x, y, \theta) &=& \prod_{n=1}^{|Q|} \bigg[\frac{1+x^2 y^2-2x y \cos( \theta - \frac{2 \pi n }{|Q|} )}{x^2+y^2-2xy \cos( \theta - \frac{2 \pi n }{|Q|} )} \bigg]^{1/6} \label{eq:h} .
\end{eqnarray}
where $ \rho_R =e^{-2\psi (R)} $. Normalization sets
\begin{eqnarray}
N &=& \int\limits_{0}^{R} \int\limits_{0}^{2 \pi} \rho(r, \theta) r dr d\theta = e^{-2\psi (R)}  h( r_d/R) R^{2}
\end{eqnarray}
where
\begin{eqnarray}
\label{eq: hx}
h(x) = \int\limits_{0}^{1}  y dy \int\limits_{0}^{2 \pi} H(x,y,\theta) d\theta \label{eq:h1} .
\end{eqnarray}
or, $\rho_R= N R^{-2}/ h(r_d/R)) $.

The total (scaled) confinement energy is 
\begin{equation}
\frac{E_{con}}{N(u_0 /\kappa^{n})} = \frac{ (\kappa R)^n}{h(r_d/R)}  \int \limits_{0}^{1} x^{n+1}dx \int\limits_{0}^{2 \pi} H(x,r_d/R, \theta)  d\theta \label{eq:EconSplit}
\end{equation} ,
and the total (scaled) interaction energy take the form
\begin{eqnarray}
\frac{ E_{int}}{N(u_0 /\kappa^{n}) } &=& \frac{\bar{N} }{2 h^2(r_d/R)} \int\limits_{0}^{1} x dx \int\limits_{0}^{2 \pi} d \theta H(x,r_d/R, \theta) \nonumber\\
& & \int\limits_{0}^{1} y dy \int\limits_{0}^{2 \pi} d \theta^{'} H(y,r_d/R,\theta') \nonumber\\
& & K_0\big[(\kappa R) \sqrt{x^2 + y^2 - 2 x y \cos(\theta - \theta^{'})}) \nonumber\\ \label{eq:K04Integral} 
\end{eqnarray}
Notable the total scaled energy $(E_{con}+E_{int})/ N(u_0 /\kappa^{n})$ is a function of the dimensionless parameters $\bar{N}$, which is fixed for a given cluster, and the variables $Q$, $\kappa R$ and $r_d/R$ which are thermodynamic degrees of freedom for the cluster.  In practice, we evaluate the function $h(r_d/R)$ by numerical integration, while the four-dimensional integral in eq (\ref{eq:K04Integral}) is evaluated using the Monte Carlo method in \textit{scikit-monaco} packages was used to calculate its value ($10^9$ sampling points are used, resulting in relative error of $\sim10^{-4}$).

\bibliography{filament2d}
\end{document}